\documentclass[a4paper,12pt,openany]{article}
\usepackage[top=20truemm,bottom=20truemm,left=23truemm,right=23truemm]{geometry}
\usepackage[]{hyperref}
\usepackage{amsmath}
\usepackage{ascmac}
\usepackage{amssymb}
\usepackage[]{graphicx}

\graphicspath{{./figs/}}

\title{D5-brane on topological black holes}
\author{Koichi Nagasaki}
\date{\today}

\begin{document}
\vspace{1cm}

\begin{center}
	{\LARGE D5-brane on topological black holes}\\
\vspace{2cm}
	{\large Koichi Nagasaki}\footnote{koichi.nagasaki24@gmail.com}\\
\vspace{1cm}
	{\small Department of Physics, 
	Toho University,\\
	Address: 2 Chome-2-1 Miyama, Funabashi, Chiba 274-8510, Japan}
\end{center}
\vspace{1.5cm}

\abstract{
Our interest is to find the difference of the behavior between black holes with three different topologies. 
These black holes have spherical, hyperbolic and toroidal structures.
We study in this paper the behavior of a probe D5-branes in this nontrivial black hole spacetime.  
We would like to find the solution what describe the embedding of a probe D5-brane.
This system realizes an ``interface" solution, a kind of non-local operators, on the boundary gauge theories.
These operators are important to deepen understanding of AdS/CFT correspondence. 
}

\vspace{1cm}
\tableofcontents
\vspace{2cm}

\section{Introduction}
The AdS/CFT correspondence \cite{Maldacena:1997re} is an important method to study the properties of the superstring theory.
If this correspondence is true, it allows one to use weak-coupling perturbative methods in one theory to study the other side of dual theory. 
However this correspondence is not proved yet.
Then we would like to find more examples and find the evidence of this conjecture. 

Generally the gravity theory has a good approximation in the region $\lambda\gg1$ while the gauge theory approximation is valid for $\lambda\ll1$, where $\lambda$ is the 't-Hooft coupling constant, $\lambda := g_\text{YM}^2N$.
Then to confirm this correspondence we have to compare the results from both theories and it is usually difficult.
To overcome this difficulty non-local operators give us the good method.
These nonlocal operators give a new parameter $k$, which appears in the result of the calculation of physical quantities in the combination of $\lambda/k^2$.
The physical quantities are expanded by the combination of $\lambda/k^2$ and by taking large $k$ limit and expanding the results in powers of $\lambda/k^2$, we can directly compare the calculations from gauge and gravity theories. 
The non-local operators are used many research so far, for example, surface operator for the AdS/CFT senanrio \cite{Constable:2002xt,Rey:2008bh,Gukov:2008sn,Drukker:2008wr} and line operators in mirror symmetry \cite{Dimofte:2019zzj}.

One of these non-local operators is an ``interface."
It separates the 4-dimensional space where gauge theories exist.
In the gravity side this non-local operator is realized by a probe D5-brane which does not affect the spacetime geometry.
The AdS spacetime is formed by multiple $N$ D3-branes. 
In this case the former parameter $k$ corresponds to the number of D3-branes ending on the D5-brane.
Then there are different number of D3-branes separated by the D5-brane and on the AdS boundary the gauge group $SU(N)$ and $SU(N-k)$ is  realized.
Our past research \cite{Nagasaki:2011ue} gives an example of these calculations.
In \cite{Nagasaki:2011ue} we considered the potential energy between an interface and a test particle which is an edge of a fundamental string ending on this interface at the other side. 
The classical gauge theory solution is fuzzy funnel solution and the potential energy between the particle and the interface is calculated as the expectation value of a Wilson loop operator. 
In gravity theory, on the other hand, the corresponding quantity is calculated from the on-shell action of the fundamental string.
We confirmed the both results agree in the first order of the expansion.
Our goal of this paper is to give the generalization of the above D5-brane solution on the black hole spacetime. 

It is worth to consider such a generalization in the following reason.
Such a holographic correspondence gives a good way to study the black hole properties and is paid attention in the study of complexity in many recent research such as
\cite{Susskind:2014rva,Brown:2015bva,Brown:2015lvg,Alishahiha:2015rta,Barbon:2015ria,Barbon:2015soa,Susskind:2015toa,Pan:2016ecg,Momeni:2016ekm,Huang:2016fks,Chapman:2016hwi,Momeni:2016ira,Reynolds:2016rvl,Carmi:2016wjl,Hashimoto:2017fga,Tao:2017fsy,Alishahiha:2017hwg,Carmi:2017jqz,Kim:2017lrw,Qaemmaqami:2017lzs,Yang:2017nfn,Karar:2017org,Abt:2017pmf,Gan:2017qkz,Gan:2017qkz,Bao:2017qmt,Miao:2017quj,Ghodrati:2017roz,Chapman:2017rqy,Guo:2017rul,HosseiniMansoori:2017tsm,Roy:2017uar,Couch:2017yil,Swingle:2017zcd,Cano:2018aqi,Hashimoto:2018bmb,An:2018dbz,Chapman:2018dem,Susskind:2018fmx,HosseiniMansoori:2018gdu,Mahapatra:2018gig,Numasawa:2018grg,Goto:2018iay,Fu:2018kcp,Chapman:2018lsv,Barbon:2018mxk,Auzzi:2018pbc,Feng:2018sqm,Jiang:2018tlu,Meng:2018vtl,An:2018xhv,Fan:2018xwf,Huang:2019ajv,Chapman:2019clq,Caginalp:2019fyt,Fan:2019mbp,An:2019opz,Jiang:2019pgc,Jiang:2019qea,Sinamuli:2019utz,Bombini:2019vuk,Karar:2019wjb,Jiang:2019yzs,Akhavan:2019zax,Bernamonti:2019zyy}.
``Complexity" is a quantity which describes how hard to create a certain state from the reference state.
It is expected that it will be a clue to solve problems about black holes, {\it e.g.}, black hole information problem and black hole firewall.
By the ``holographic complexity" conjecture complexity is expected to be dual to the bulk object --- action or volume.
Then it is worthwhile to generalize the previous interface method to the black hole spacetime in order to use it in the study of holographic complexity.

Our goal of this paper is to generalize the interface solution $x_3=\kappa y$ found in \cite{Nagasaki:2011ue} to the black hole spacetime, especially a black hole whose event horizon has surfaces of non-trivial topology.
These black holes are known as topological black holes \cite{Lemos:1995cm,Aminneborg:1996iz,Mann:1997iz,Vanzo:1997gw}, especially the planer case ($k=0$) is studied in detail in \cite{Lemos:1994fn,Lemos:1994xp,Cai:1996eg}.
Such black holes also can have the angular momentum \cite{Klemm:1997ea} and can be an interesting research theme as well as the usual spherical black holes.
For example, thermodynamics of these black holes is studied in \cite{Brill:1997mf}.
Such a spacetime has the asymptotic AdS structure \cite{Birmingham:1998nr,10.1093/ptep/pty017} and then to be very important for finding a new example of the AdS/CFT dualities.
These solutions are distinguished by the horizon geometry of a $(d-2)$-dimensional Einstein manifold which has positive, zero, or negative curvature.

The construction of this paper is as follows: 
We first study the case of static case in Section \ref{sec:static_topBHs}.
These black holes are with positive, negative and zero curvatures.
Such three different topologies are distinguished by parameter $k$.
In the above we used $k$ for counting the number of D3-branes or gauge groups but it is not confusing since we use parameter $k=\pm1$ to distinguish these three topologies for the most part of this paper except the last section where we discussed boundary gauge theories.
In Section \ref{sec:rot_topBHs} we study more general case where black holes with an angular momentum.
In this case these three kind of topology show very different behavior.
This is what we want to show in this paper.

\section{Static topological black hole cases}\label{sec:static_topBHs}
\subsection{Static topological black holes}
The metric of the asymptotically AdS black hole spacetime 
\begin{equation}
ds_{k,d+1}^2 = -f_k(r)dt^2 + \frac{dr^2}{f_k(r)} + r^2d\Sigma_{k,d-1}^2,
\end{equation}
with the angular part
\begin{equation}
d\Sigma_{k,d-1}^2
= \begin{cases}
d\theta^2 + \sin^2\theta d\Omega_{d-2}^2, & (k=+1),\\
d\theta^2 + \sinh^2\theta d\Omega_{d-2}^2, & (k=-1),\\
d\theta^2 + d\Omega_{d-2}^2, & (k=0),
\end{cases}
\end{equation}
and the metric function $f_k(r)$ is 
\begin{equation}
f_k(r) = k + r^2 - \frac{r_\text{m}^{d-2}}{r^{d-2}},\;\;
r_\text{m}^{d-2} = \frac{16\pi G_{d+1}M}{(d-1)\Sigma_{k,d-1}}.
\end{equation}
We define a function $s_k(\theta)$ such that 
\begin{equation}\label{eq:staticmetric_sk}
s_k(\theta) = \begin{cases}
\sin^2\theta & (k=+1),\\ \sinh^2\theta & (k=-1),\\ 1 & (k=0),
\end{cases}
\end{equation}
where the variable $\theta$ varies in the range
$0\leq\theta\leq\pi$ for spherical and the toroidal cases and periodically identified for the toroidal case. 
By this notation, we write the angular part as 
$d\Sigma_{k,d-1} = d\theta^2 + s_k(\theta)d\Omega_{d-2}^2$ in the following.

We choose the coordinates as 
$(t,r,\theta,\phi,\psi)$ on $\text{AdS}_5$ and $(\varphi,\chi,\chi_1,\chi_2,\chi_3)$ on $S^5$.
We consider a D5-brane which extends to the directions
$t$, $\phi$, $\psi$ in $\text{AdS}_5$ and $\varphi$ and $\chi$ in $S^5$ planarly and appears a one-dimensional object in $(r,\theta)$ plane.
It is convenient to choose the radial coordinates $y=1/r$ where $y=0$ corresponds to the infinity. 
In these coordinates, the metric is 
\begin{equation}
ds_{k,5}^2|_\text{ind} 
= \frac1{y^2}\Big(-F_k(y)dt^2 + \frac{dy^2}{F_k(y)} + d\Sigma_{k,3}^2\Big),
\end{equation}
where we defined a new metric function as
\begin{equation}
F_k(y) 
:= y^2f_k(y^2) 
= 1 + ky^2 - r_\text{m}^{d-2}y^d.
\end{equation}
It goes to zero in the limit $y\rightarrow0$. 
Then it certainly recovers the flat AdS spacetime sufficiently far from black holes.

There is a gauge flux on the D5-brane we assume it extends on the $S^2$ subspace on $S^5$.
Then our assumption is 
\begin{equation}
\mathcal F = -\kappa\; d\text{vol}[S^2]
= -\kappa\sin\varphi d\varphi\wedge d\chi.
\end{equation}
Since the D5-brane describes one-dimensional subspace on $(y,\theta)$-plane, it is written by one parameter expression:
\begin{equation}
y = y(\sigma),\;\;
\theta = \theta(\sigma).
\end{equation}

By this assumption, the induced metric for the AdS part is
\begin{equation}
ds_\text{AdS}^2|_\text{ind} 
= \frac1{y^2}\Big(-F_k(y)dt^2 + \Big(\frac{y'^2}{F_k(y)} + \theta'^2\Big)d\sigma^2 + s_k(\theta)(d\phi^2 + \sin^2\phi d\psi^2)\Big).
\end{equation}
The sphere direction is 
\begin{equation}
ds_{S^5}^2|_\text{ind}
= d\varphi^2 + \sin^2\varphi d\chi^2.
\end{equation}
The summation of the metric and the flux is,
\begin{equation}
(g_k + \mathcal F)|_\text{ind}
= \frac1{y^2}\text{diag}\Big[
  -F_k,\frac{y'^2}{F_k} + \theta'^2,s_k,s_k\sin^2\phi\Big]\oplus
\begin{bmatrix}
1& -\kappa\sin\varphi\\ \kappa\sin\varphi& \sin^2\varphi
\end{bmatrix}.
\end{equation}

\subsection{D5-brane action and equation of motions}
The D5-brane action consist of the DBI part and the Wess-Zumino part: 
\begin{subequations}
\begin{align}
S &= S_\text{DBI} + S_\text{WZ},\\
S_\text{DBI} &= -T_5\int\sqrt{-\det(g_k+\mathcal F)_\text{ind}},\\
S_\text{WZ} &= T_5\int\mathcal F\wedge C_4.
\end{align}
\end{subequations}

The DBI action is by the calculation of the determinant
\begin{equation}
\sqrt{-\det(g_k+\mathcal F)_\text{ind}}
= \frac1{y^4}\sqrt{y'^2 + \theta'^2F_k}\; s_k\sin\phi\sqrt{1+\kappa^2}\sin\varphi. 
\end{equation}

Let us compute the Wess-Zumino part. 
In this expression $C_4$ is the Ramond-Ramond 4-form which satisfies
$dC_4 = 4d\text{vol[AdS]} + 4d\text{vol}[S^5]$.
The volume form of the AdS spacetime is
\begin{align*}
d\text{vol[AdS]} = -\frac1{y^5}s_k\sin\phi dtdyd\theta d\phi d\psi.
\end{align*}
We find the $C_4$ as
\begin{equation}
C_4 = -\frac1{y^4}s_k\sin\phi dtd\theta d\phi d\psi + 4\alpha_4,
\end{equation}
where $\alpha_4$ is the 4-form on the $S^5$ which satisfies $d\alpha_4 = d\text{vol}[S^5]$.

Taking the product with gauge flux $\mathcal F\wedge C_4$, $\alpha_4$ part vanishes and gives
\begin{equation}
\mathcal F\wedge C_4 
= \frac{\kappa}{y^4}s_k\sin\phi\sin\varphi dtd\theta d\phi d\psi d\varphi d\chi,
\end{equation}
where the first sine is from the AdS and the second one is from the sphere part. 

The D5-brane action is
\begin{align}
S &= -T_5\int_\text{AdS}dtd\sigma d\phi d\psi\int_{S^2}d\varphi d\chi
 \frac{s_k\sin\phi\sin\varphi}{y^4}
 \Big(\sqrt{1+\kappa^2}\sqrt{y'^2 + \theta'^2F_k} - \kappa\theta'\Big)\nonumber\\
&= -(4\pi)^2T_5\Delta t\;\int d\sigma
 \frac{s_k}{y^4}\Big(\sqrt{1+\kappa^2}\sqrt{y'^2 + \theta'^2F_k} - \kappa\theta'\Big).
\end{align}
In the above we performed the integral in the angular directions.
The Lagrangian is, eliminating the whole constant factor,
\begin{equation}
\mathcal L(\sigma) 
= \frac{s_k(\theta)}{y^4}
  \big(\sqrt{y'^2 + \theta'^2F_k} - \tilde\kappa\theta'\big),
\end{equation}
where we defined $\tilde\kappa := \kappa(1+\kappa^2)^{-1/2}$.

The derivative by each variable is as follows.
For $y$,
\begin{subequations}
\begin{align}
\frac{\partial\mathcal L}{\partial y'}
&= \frac{s_k}{y^4}\frac{y'}{\sqrt{y'^2+\theta'^2F_k}},\\
\frac{\partial\mathcal L}{\partial y}
&= -4\frac{s_k}{y^5}(\sqrt{y'^2 + \theta'^2F_k} - \tilde\kappa\theta')
 + \frac{s_k}{y^4}\frac{\theta'^2\partial F_k}{2\sqrt{y'^2+\theta'^2F_k}}.
\end{align}
\end{subequations}
For $\theta$,
\begin{subequations}
\begin{align}
\frac{\partial\mathcal L}{\partial\theta'}
&= \frac{s_k}{y^4}\Big(\frac{\theta'F_k}{\sqrt{y'^2+\theta'^2F_k}} - \tilde\kappa\Big),\\
\frac{\partial\mathcal L}{\partial\theta}
&= \frac{\partial s_k}{y^4}
  \big(\sqrt{y'^2 + \theta'^2F_k} - \tilde\kappa\theta'\big).
\end{align}
\end{subequations}
In the above expressions by $\partial$ we mean the derivative by $y$ for $F_k(y)$ and that by $\theta$ for $s_k(\theta)$.

Let us fix the gauge freedom as 
$\sqrt{y'^2 + \theta'^2F_k} = 1$.
The above equations are simplified as
\begin{subequations}
\begin{align}
\frac{\partial\mathcal L}{\partial y'}
&= y'\frac{s_k}{y^4},\\
\frac{\partial\mathcal L}{\partial y}
&= -4(1-\tilde\kappa\theta')\frac{s_k}{y^5}
 + \frac12\theta'^2\partial F_k\frac{s_k}{y^4},\\
\frac{\partial\mathcal L}{\partial\theta'}
&= (\theta'F_k - \tilde\kappa)\frac{s_k}{y^4},\\
\frac{\partial\mathcal L}{\partial\theta}
&= (1-\tilde\kappa\theta')\frac{\partial s_k}{y^4}.
\end{align}
\end{subequations}
The equations of motion are
\begin{subequations}
\begin{align}
\frac{d}{d\sigma}\Big(y'\frac{s_k}{y^4}\Big)
 + 4(1-\tilde\kappa\theta')\frac{s_k}{y^5}
 - \frac12\theta'^2\partial F_k\frac{s_k}{y^4} = 0,\\
\frac{d}{d\sigma}\Big((\theta'F_k - \tilde\kappa)\frac{s_k}{y^4}\Big)
 - (1-\tilde\kappa\theta')\frac{\partial s_k}{y^4} = 0.
\end{align}
\end{subequations}
By dividing by the common factor, the equations are
\begin{subequations}
\begin{align}
y'' + y'\frac{d}{d\sigma}\log\frac{s_k}{y^4}
 + \frac4y(1-\tilde\kappa\theta')
 - \frac12\theta'^2\partial F_k = 0,\\
\theta''F_k + \theta'y'\partial F_k
 + (\theta'F_k - \tilde\kappa)\frac{d}{d\sigma}\log\frac{s_k}{y^4}
 - (1 - \tilde\kappa\theta')\partial\log s_k = 0.
\end{align}
\end{subequations}

We fixed the gauge $y'^2 + \theta'^2F_k = 1$.
Let us confirm it is preserved along the solution.
By differentiating by $\sigma$, 
\begin{equation}
\frac{d}{d\sigma}(y'^2 + \theta'^2F_k)
= 2y'y'' + 2\theta'\theta''F_k + \theta'^2y'\partial F_k.
\end{equation}
Substituting the equation of motion,
\begin{align}
\frac{d}{d\sigma}(y'^2 + \theta'^2F_k)
&= 2y'\Big(-y'\frac{d}{d\sigma}\log\frac{s_k}{y^4} 
  - \frac4y(1-\tilde\kappa\theta')
  + \frac12\theta'^2\partial F_k\Big)\nonumber\\
&\quad
 + 2\theta'\Big(-\theta'y'\partial F_k 
  - (\theta'F_k-\tilde\kappa)\frac{d}{d\sigma}\log\frac{s_k}{y^4}
  + (1-\tilde\kappa\theta')\partial\log s_k\Big)
 + \theta'^2y'\partial F_k\nonumber\\
&= -2(y'^2 + \theta'^2F_k - \tilde\kappa\theta')\frac{d}{d\sigma}\log\frac{s_k}{y^4}
 + 2(1-\tilde\kappa\theta')\frac{d}{d\sigma}\log(y^{-4})
 + 2(1 - \tilde\kappa\theta')\frac{d}{d\sigma}\log s_k\nonumber\\
&= -2(y'^2 + \theta'^2F_k - 1)\frac{d}{d\sigma}\log\frac{s_k}{y^4},
\end{align}
where the last expression is zero if we use the gauge condition. 
By this calculation we confirmed that the gauge condition $y'^2+\theta'^2F_k = 1$ is satisfied along the surface.

The equations are summarized as
\begin{subequations}\label{eq:topBH_static_eom}
\begin{align}
y'' + y'\theta'\partial\log s_k
 + \frac4y(1-y'^2-\tilde\kappa\theta')
 - \frac12\theta'^2\partial F_k = 0,\\
\theta''F_k + \theta'y'\partial F_k
 + (\theta'^2F_k - 1)\partial\log s_k
 - \frac4y(y'\theta'F_k - \tilde\kappa y') = 0.
\end{align}
\end{subequations}

\subsection{Behavior in specific regions}
\begin{figure}[h]
\begin{center}
	\includegraphics[width=12cm]{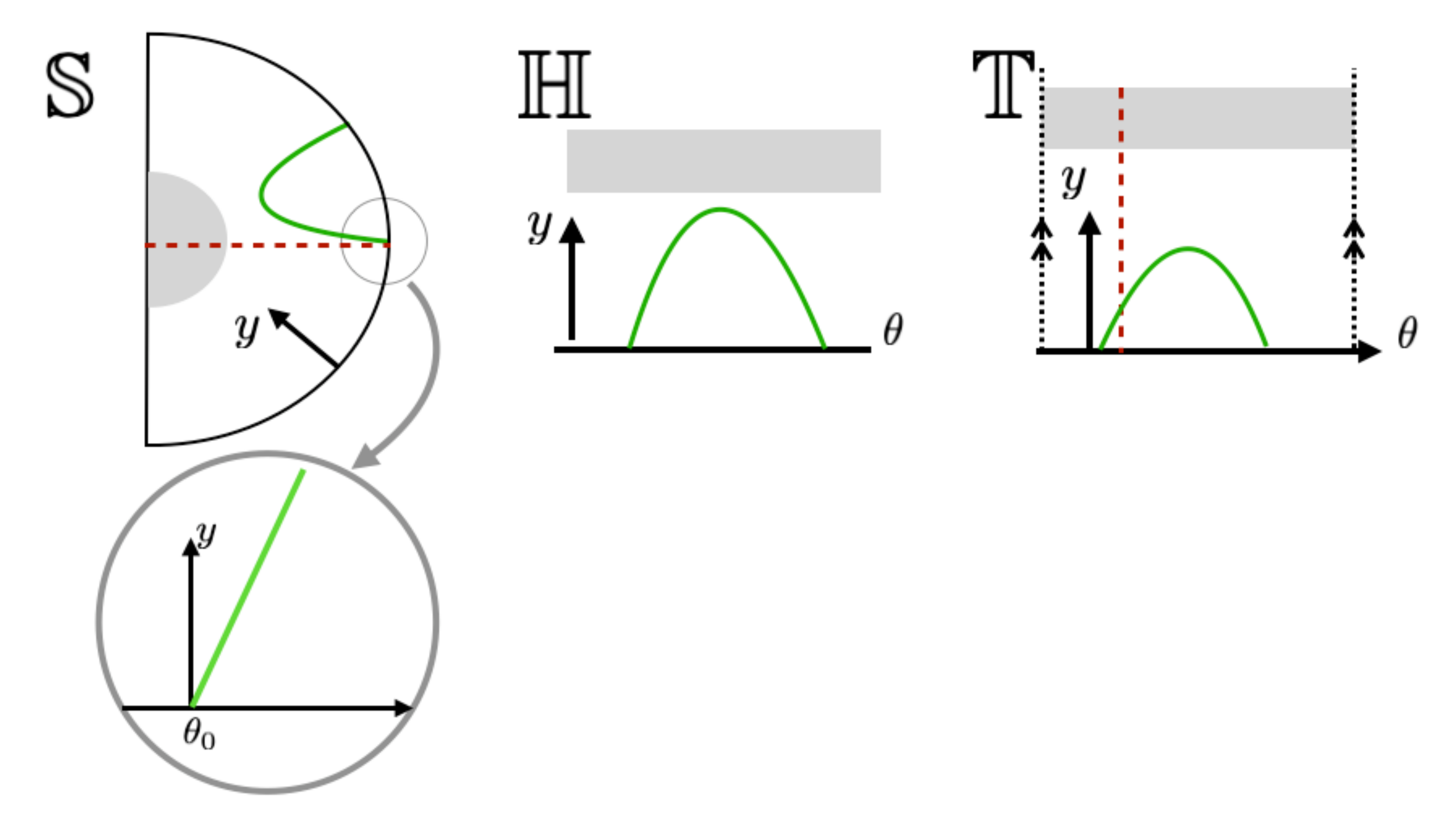}
	\caption{D5-brane in AdS black hole spacetime}
	\label{fig:D5_in_topBH}
\end{center}
\end{figure}

\paragraph{No flux case}
We can see that when $\tilde\kappa=0$, $y = \sigma$ and $\theta = \text{const}$ can be a solution for the first equation and the second equation is satisfied as long as 
$\partial\log s_k = 0$.
The function $\partial\log s_k$ is $2\cot\theta$, $2\coth\theta$ or $0$ for $\mathbb S$, $\mathbb H$ and $\mathbb T$ cases, respectively (see Eq.\eqref{eq:staticmetric_sk}). 
This condition is always satisfied for $\mathbb T$.
For $\mathbb S$ constant $\theta$ must be fixed to $\theta_0 = \pi/2$ and for $\mathbb H$ there is no such a solution ($\coth x > 1$).
Such solutions are depicted by red dashed lines in Figure \ref{fig:D5_in_topBH}.

\paragraph{Boundary behavior}
As go to the infinity, the metric function approaches $F_k\rightarrow1$ and 
$\partial F_k = 2ky - dr_\text{m}^{d-2}y^{d-1}
\rightarrow0$ ($d=4$) for all cases ($\mathbb S$, $\mathbb H$ and $\mathbb T$).
Since this spacetime approaches flat AdS at infinity, the solution must approach the one on the flat AdS space shown in \cite{Nagasaki:2011ue}.
It was $x_3 = \kappa y$, where $x_3$ is a usual Cartesian coordinate and $y$ is the radial direction.
In our coordinates it is 
$\kappa y = x_3 = X_3y = (r\cos\theta)y = \cos\theta$.
\footnote{
In coordinates $X_i; (i=0,1,\cdots5)$, the metric is 
\begin{equation}
ds^2 = - dX_0^2 - dX_5^2 + \sum_{i=1}^4 dX_i^2,\quad
-X_0^2-X_5^2+\sum_{i=1}^4 X_i^2 = -\ell_\text{AdS}^2.
\end{equation}
By the coordinate transformation,
\begin{equation}
X_0 = \frac1{2y}(x^2 + y^2 + 1),\;\;
X_5 = \frac{x^0}{y},\;\;
X_i = \frac{x^i}{y};\; (i=1,2,3),\;\;
X_4 = \frac1{2y}(x^2 + y^2 - 1),
\end{equation}
this metric becomes 
\begin{equation}
ds^2 = \frac{dy^2 + \eta_{\mu\nu}dx^\mu dx^\nu}{y^2}.
\end{equation}
}
That is $\cos\theta = \kappa y$.
Expanding near $\theta_0 = \pi/2$ gives
$\kappa y = \cos(\pi/2-\delta\theta)\sim\sin\delta\theta
\sim\delta\theta$ and 
$\delta\theta\sim\kappa y$.
Then we have
$\delta\theta'(\sigma)\sim\kappa y'(\sigma)$.
The gauge condition becomes 
$1 = y'^2 + \delta\theta'^2F_k 
= (1+\kappa^2)y'^2$.
By solving it for $y'$, 
\begin{equation}
y'(\sigma=0) = \frac1{\sqrt{1+\kappa^2}},\;\;
\delta\theta'(\sigma=0) = \frac{\kappa}{\sqrt{1+\kappa^2}}.
\end{equation}
We impose the boundary condition where $y(\sigma=0) = 0$, $\theta(\sigma=0) = \pi/2$ and their first derivatives are given above. 

\subsection{Numerical results}
\begin{figure}[h]
	\begin{minipage}[t]{0.5\linewidth}
	\includegraphics[width=\linewidth]{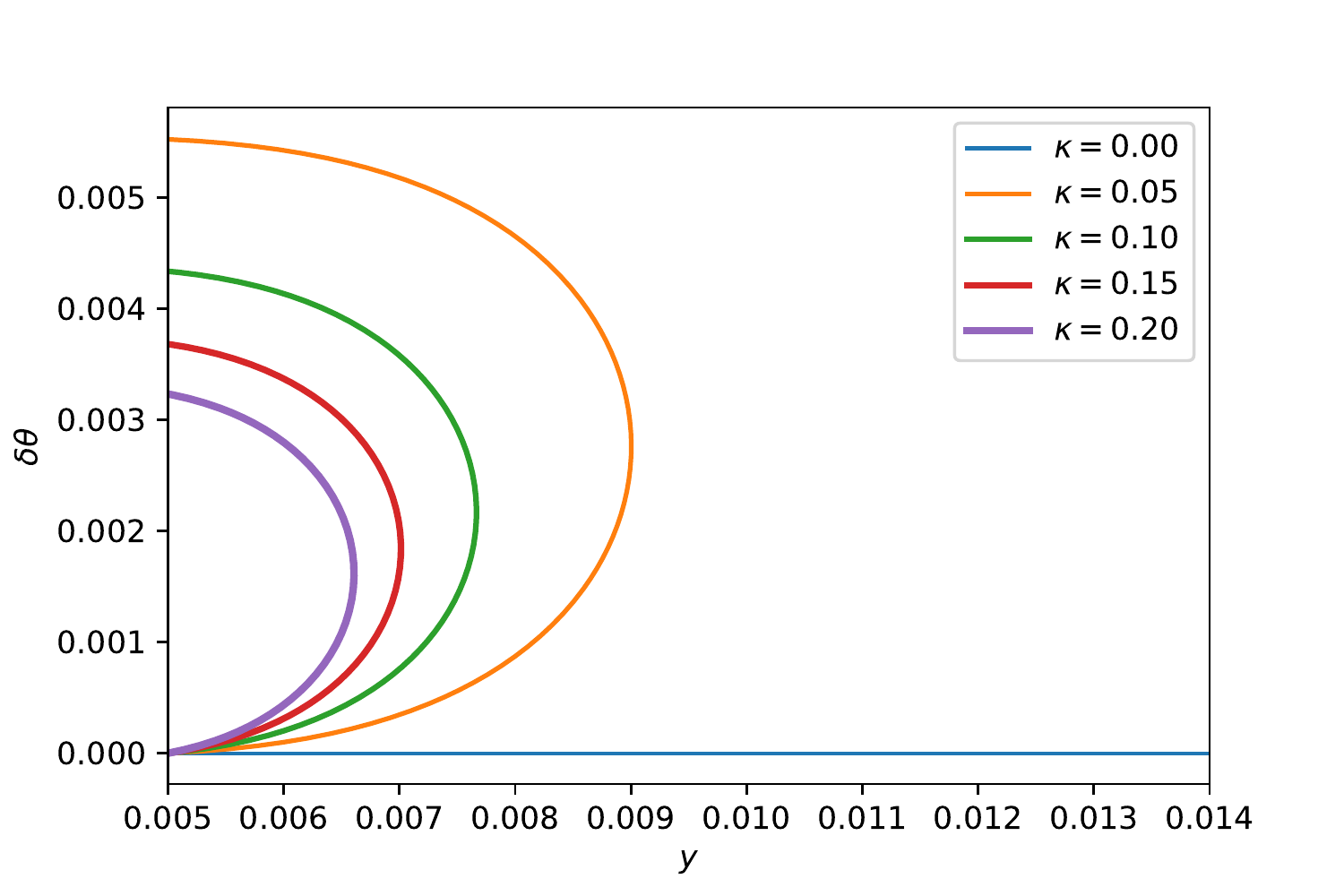}
	\caption{Sphere case with different fluxes}
	\label{fig:topBH_static_sphere}
	\end{minipage}
\hspace{0.01\linewidth}
	\begin{minipage}[t]{0.5\linewidth}
	\includegraphics[width=\linewidth]{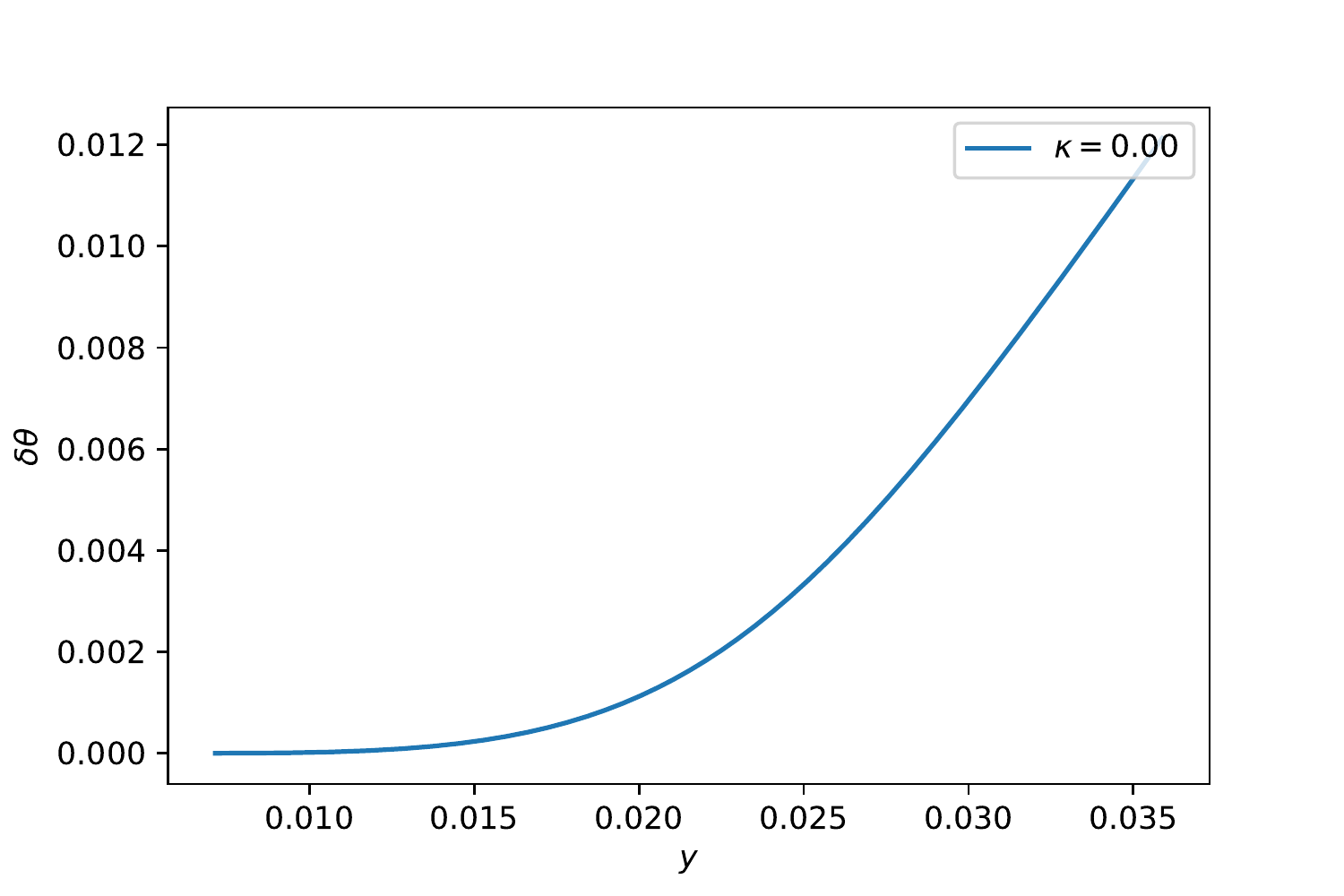}
	\caption{Hyperbolic case}
	\label{fig:topBH_static_hyperbolic0}
	\end{minipage}
\end{figure}

\begin{figure}[h]
	\begin{minipage}[t]{0.5\linewidth}
	\includegraphics[width=\linewidth]{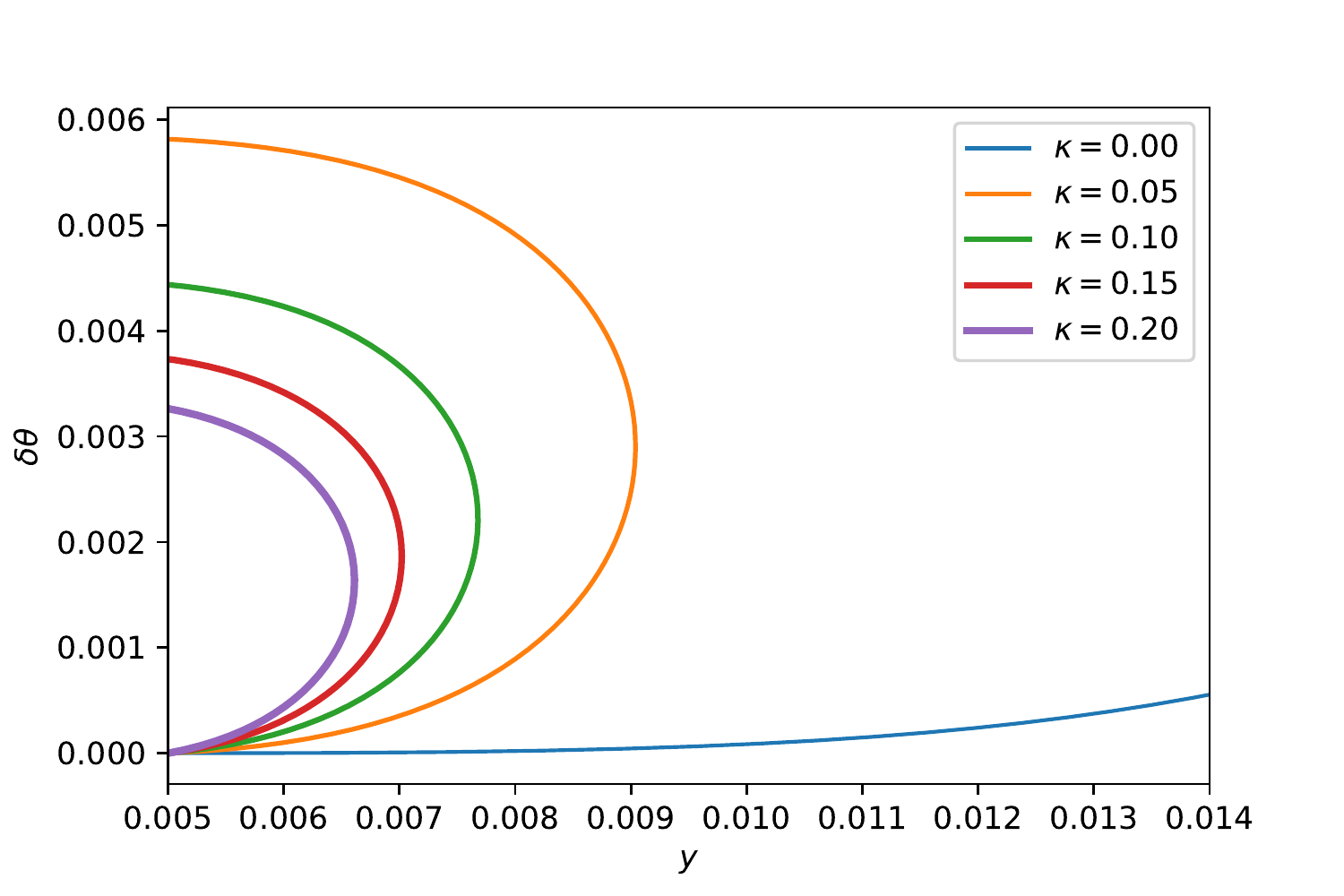}
	\caption{Hyperbolic case}
	\label{fig:topBH_static_hyperbolic}
	\end{minipage}
\hspace{0.01\linewidth}
	\begin{minipage}[t]{0.5\linewidth}
	\includegraphics[width=\linewidth]{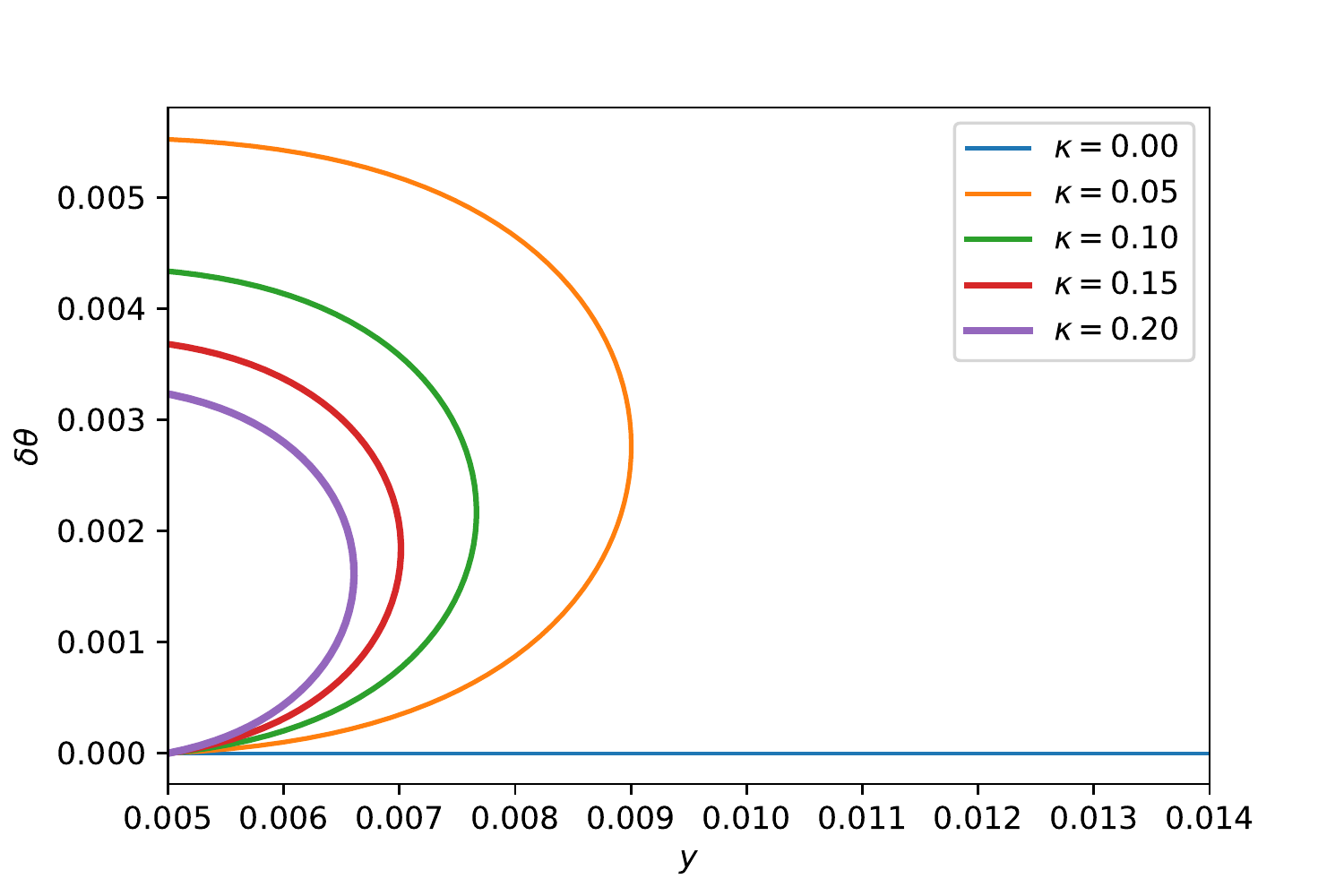}
	\caption{Toroidal case with different fluxes}
	\label{fig:topBH_static_torus}
	\end{minipage}
\end{figure}

The results of the numerical calculation is shown in Figures \ref{fig:topBH_static_sphere}, \ref{fig:topBH_static_hyperbolic} and \ref{fig:topBH_static_torus}.
In this plot the vertical axis $\delta\theta$ is the displacement from the trivial solution: 
$\delta\theta = \theta - \pi/2$.
We summarize the main result.
\begin{enumerate}
\item The trivial solution $\theta = \text{constant}$ can exist for spherical and the torus cases and for hyperbolic case there is no such a solution. 
\item For sphere case the above constant must be $\pi/2$ and for torus case this constant can take arbitrary values between $0\leq\theta\leq\pi$ ($\theta$ is periodically identified with periodicity $\pi$).
\end{enumerate}

\newpage
\section{Rotating topological black hole cases}\label{sec:rot_topBHs}
\subsection{Three kinds of topology}
The rotating topological black hole solution is given in \cite{Klemm:1997ea}.
Before the detailed analysis, let us review these solutions.
We use the coordinates where the metric for $S^3$ subspace is 
$ds^2 = d\theta^2 + \sin^2\theta d\phi_1^2 + \cos^2\theta d\phi_2^2$.
For this coordinate system the angular coordinate takes the value in $\theta\in[0,\pi/2]$.

For $k=1$, the black hole has the spherical structure (we denote it as $\mathbb S$ below)
\begin{align}
ds_{\text{KA}\mathbb S}^2
&= \frac{\rho^2}{\Delta_r}dr^2
 + \frac{\rho^2}{\Delta_\theta}d\theta^2
 - \frac{\Delta_r}{\rho^2\Xi^2}
  (dt - a\sin^2\theta d\phi)^2\nonumber\\
&\qquad
 + \frac{\Delta_\theta\sin^2\theta}{\rho^2\Xi^2}
  (adt - (r^2+a^2)d\phi)^2
  + r^2\cos^2\theta d\Omega_{1,d-3}^2,
\end{align}
where
\begin{align}
\Delta_r &:= (r^2+a^2)(r^2+1) - \frac{2m}{r^{d-4}},\nonumber\\
\Delta_\theta &:= 1 - a^2\cos^2\theta,\nonumber\\
\rho^2 &:= r^2 + a^2\cos^2\theta,\qquad
\Xi := 1 - a^2.
\end{align}
For $k= -1$, the black hole has the hyperbolic space structure (we denote it as $\mathbb H$ below)
\begin{align}
ds_{\text{KA}\mathbb H}^2
&= \frac{\rho^2}{\Delta_r}dr^2
 + \frac{\rho^2}{\Delta_\theta}d\theta^2
 - \frac{\Delta_r}{\rho^2\Xi^2}
  (dt + a\sinh^2\theta d\phi)^2\nonumber\\
&\qquad
 + \frac{\Delta_\theta\sinh^2\theta}{\rho^2\Xi^2}
  (adt - (r^2+a^2)d\phi)^2
  + r^2\cosh^2\theta d\Omega_{-1,d-3}^2,
\end{align}
where
\begin{align}
\Delta_r &:= (r^2+a^2)(r^2-1) - \frac{2m}{r^{d-4}},\nonumber\\
\Delta_\theta &:= 1 + a^2\cosh^2\theta,\nonumber\\
\rho^2 &:= r^2 + a^2\cosh^2\theta,\qquad
\Xi := 1 + a^2.
\end{align}
For $k= 0$, the metric of the toroidal rotating black hole (we denote it as $\mathbb T$) is
\begin{equation}
ds_{\text{KA}\mathbb T}^2
= - N^2dt^2 
  + \frac{\rho^2}{\Delta_r}dr^2
  + \frac{\rho^2}{\Delta_\theta}d\theta^2
  + \frac{\Sigma^2}{\rho^2}(d\phi - \omega dt)^2
  + r^2d\Omega_{0,d-3}^2,
\end{equation}
where
\begin{align}
\Delta_r &:= r^4 + a^2 - \frac{2m}{r^{d-4}},\nonumber\\
\Delta_\theta &:= 1 + a^2\theta^4,\nonumber\\
\rho^2 &:= r^2 + a^2\theta^2,\qquad
\Sigma^2 := r^4\Delta_\theta - a^2\theta^4\Delta_r,\nonumber\\
\omega &:= a\frac{\Delta_r\theta^2 + r^2\Delta_\theta}{\Sigma^2},\qquad
N^2 := \frac{\rho^2\Delta_\theta\Delta_r}{\Sigma^2}.\nonumber
\end{align}
The time and angular parts of the metric are
\begin{equation}
-N^2dt^2 + \frac{\Sigma^2}{\rho^2}(d\phi-\omega dt)^2
= \Big(-\frac{\rho^2\Delta_r\Delta_\theta}{\Sigma^2}
  + \frac{\Sigma^2\omega^2}{\rho^2}\Big)dt^2
 + \frac{\Sigma^2}{\rho^2}d\phi^2
 - 2\frac{\Sigma^2}{\rho^2}\omega dtd\phi.
\end{equation}
The coefficient of $dt^2$ is 
\begin{align}
-\frac{\rho^2\Delta_r\Delta_\theta}{\Sigma^2}
  + \frac{\Sigma^2\omega^2}{\rho^2}
= \frac1{\rho^2\Sigma^2}
  (\Delta_r-a^2\Delta_\theta)(a^2\theta^4\Delta_r-r^4\Delta_\theta)
= -\frac{\Delta_r-a^2\Delta_\theta}{\rho^2}.
\end{align}
Then $dt$ and $d\phi$ parts are summarized as
\begin{align}
&\rho^{-2}(-\Delta_rdt^2 + a^2\Delta_\theta dt^2
 - 2(a\theta^2\Delta_r + ar^2\Delta_\theta)dtd\phi
 - (a^2\theta^4\Delta_r-r^4\Delta_\theta)d\phi^2)\nonumber\\
&= \rho^{-2}[-\Delta_r(dt + a\theta^2d\phi)^2 + \Delta_\theta(adt - r^2d\phi)^2].
\end{align}
Therefore the metric for torus case can be shown in similar form to the spherical and hyperbolic cases, 
\begin{subequations}
\begin{align}
ds_{\text{KA}\mathbb T}^2
&= \frac{\rho^2}{\Delta_r}dr^2 + \frac{\rho^2}{\Delta_\theta}d\theta^2
 - \frac{\Delta_r}{\rho^2}(dt + a\theta^2d\phi)^2
 + \frac{\Delta_\theta}{\rho^2}(adt - r^2d\phi)^2
 + r^2d\Omega_{0,d-3}^2,\\
\Delta_r &= r^4+a^2-\frac{2m}{r^{d-4}},\;\;
\Delta_\theta = 1+a^2\theta^4,\;\;
\rho^2 = r^2 + a^2\theta^2.
\end{align}
\end{subequations}

To write the above three cases uniformly we define a function $s_k(\theta)$ such that 
\begin{equation}\label{eq:topBH_rot_sk}
s_k(\theta) = \begin{cases}
\sin^2\theta & (k=1),\\ \sinh^2\theta & (k=-1),\\ 1 & (k=0),
\end{cases}
\end{equation}
and also define a function
\begin{equation}
t_k(\theta) := (k^2-1)\theta^2 + ks_k(\theta).
\end{equation}
By these functions the metrics for three cases are summarized as 
\begin{align}
ds_{k,d+1}^2
&= -\frac{\Delta_r}{\rho^2\Xi^2}(dt - at_k(\theta)d\phi)^2
 + \rho^2\Big(\frac{dr^2}{\Delta_r} + \frac{d\theta^2}{\Delta_\theta}\Big)\nonumber\\
&\qquad
 + \frac{\Delta_\theta s_k(\theta)}{\rho^2\Xi^2}
  (adt - (r^2+k^2a^2)d\phi)^2
 + r^2(1-ks_k(\theta))d\Omega_{d-3}^2.
\end{align}
The functions in this metric are 
\begin{subequations}
\begin{align}
\Delta_r &= r^4 + (1+ka^2)kr^2 + (1+k-k^2)a^2 - \frac{2m}{r^{d-4}},\\
\Delta_\theta &= 1 - a^2(k-k^2s_k-(1-k^2)\theta^4),\\
\rho^2 &= r^2 + a^2(k^2-ks_k+(1-k^2)\theta^2),\;\;
\Xi = 1-ka^2.
\end{align}
\end{subequations}

\subsection{$d=4$ case}
\begin{figure}[h]
\begin{center}
	\includegraphics[width=8cm]{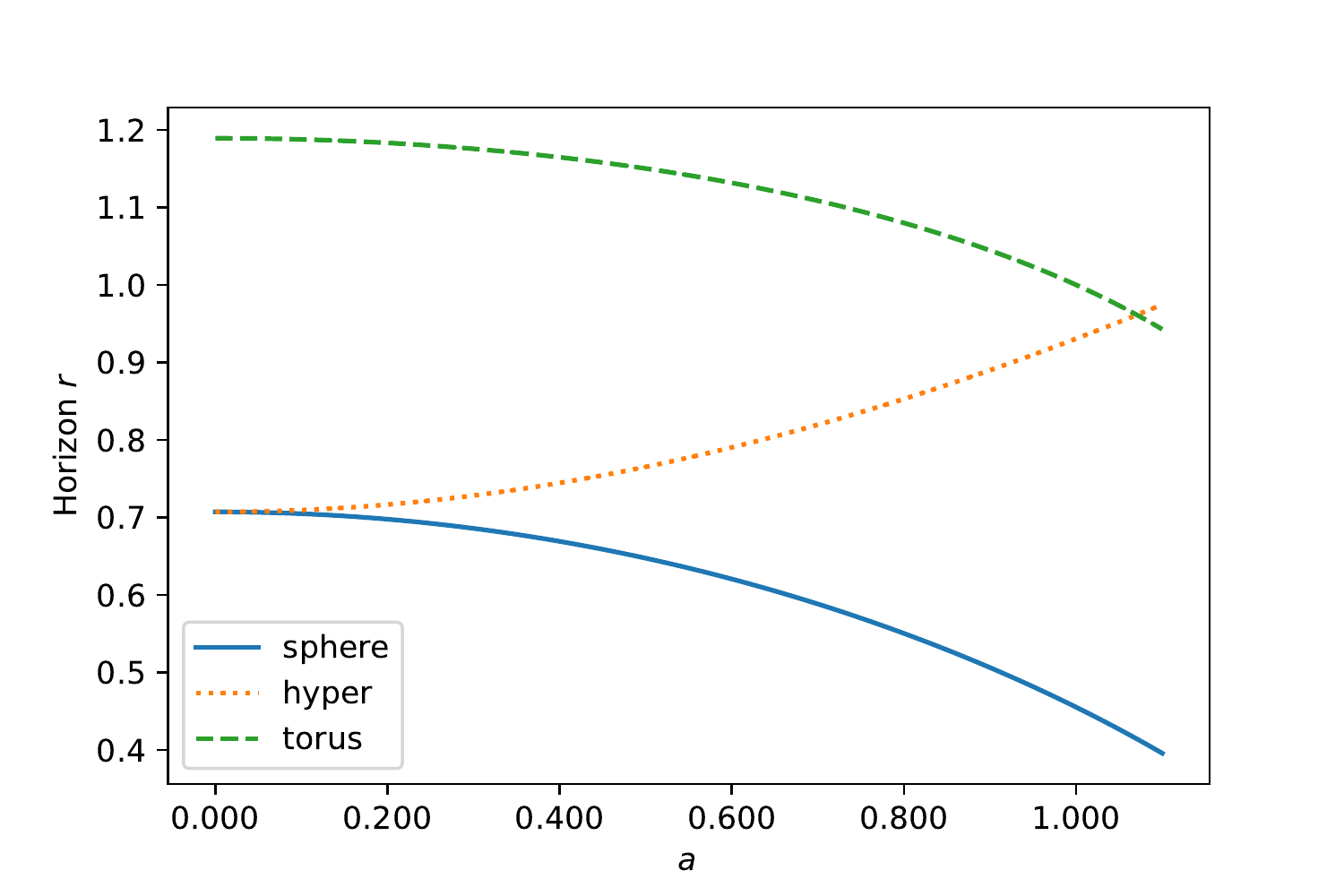}
	\caption{Angular momentum dependence of the horizon for three cases}
	\label{fig:topBH_rot_horizon}
\end{center}
\end{figure}

For $d=4$ case, it is explicitly written as
\begin{align}
ds_{k,5}^2
&= -\frac{\Delta_r}{\rho^2\Xi^2}
  (dt - at_k(\theta)d\phi)^2
 + \rho^2\Big(\frac{dr^2}{\Delta_r} + \frac{d\theta^2}{\Delta_\theta}\Big)\nonumber\\
&\qquad
 + \frac{\Delta_\theta}{\rho^2\Xi^2}s_k(\theta)
  (adt - (r^2+a^2)d\phi)^2
 + r^2(1-ks_k(\theta))d\psi^2,
\end{align}
where the metric of the sphere part is one-dimensional, that is 
$d\Sigma_{k,1}^2 = d\psi^2$.
We would like to change the coordinates as before, $y=1/r$,
\begin{align}
ds_{k,5}^2
&= -\frac{\Delta_r}{\rho^2\Xi^2}
  (dt - at_k(\theta)d\phi)^2
 + \rho^2\Big(\frac{dy^2}{y^4\Delta_r} + \frac{d\theta^2}{\Delta_\theta}\Big)\nonumber\\
&\qquad
 + \frac{\Delta_\theta}{\rho^2\Xi^2}s_k(\theta)
  \Big(adt - \Big(\frac1{y^2}+a^2\Big)d\phi\Big)^2
 + \frac1{y^2}(1-ks_k(\theta))d\psi^2,
\end{align}
and define $y^4\Delta_r =: \Delta_y$ and $y^2\rho^2 =: \rho_y^2$.
By this definition, 
\begin{align}
ds_{k,5}^2 
&= \frac1{y^2}\Big[-\frac{\Delta_y}{\rho_y^2\Xi^2}(dt - at_kd\phi)^2
 + \rho_y^2\Big(\frac{dy^2}{\Delta_y} + \frac{d\theta^2}{\Delta_\theta}\Big)\nonumber\\
&\qquad
 + \frac{\Delta_\theta}{\rho_y^2\Xi^2}s_k(ay^2dt - (1+a^2y^2)d\phi)^2
 + (1-ks_k)d\psi^2\Big].
\end{align}
In the above the functions in the metric are
\begin{subequations}
\begin{align}
\Delta_y &= 1 + (1+ka^2)ky^2 + (1+k-k^2)a^2y^4 - 2my^d,\\
\Delta_\theta &= 1 - a^2(k-k^2s_k-(1-k^2)\theta^4),\\
\rho_y^2 &= 1 + (k^2-ks_k+(1-k^2)\theta^2)a^2y^2,\qquad
\Xi = 1 - ka^2.
\end{align}
\end{subequations}
The horizon determined by $\Delta_y=0$ is described in Figure \ref{fig:topBH_rot_horizon}.
For sphere and torus cases, the angular momentum decreases the horizon radius while the angular momentum increases the horizon radius for hyperbolic case.

The volume form is 
\begin{align}
&-\frac{\sqrt{s_k(1-ks_k)}}{y^5\Xi^2}(dt - at_k d\phi)
 dyd\theta(ay^2dt - (1+a^2y^2)d\phi)d\psi\nonumber\\
&= \frac{\sqrt{s_k(1-ks_k)}}{\Xi^2}
  \Big(\frac1{y^5} + \frac{a^2}{y^3}(1+t_k)\Big)dtd\phi dyd\theta d\psi.
\end{align}
The RR4-form, $dC_4 = 4d(\text{vol})$, is 
\begin{equation}\label{eq:topBH_rot_RR4form}
C_4 = -\frac{\sqrt{s_k(1-ks_k)}}{\Xi^2}
  \Big(\frac1{y^4} + \frac{2a^2}{y^2}(1+t_k)\Big)dtd\phi d\theta d\psi.
\end{equation}
By the same assumption as before, the induced metric is 
\begin{align}
ds_{k,5}^2|_\text{ind} 
&= \frac1{y^2}\Big[
 -\frac{\Delta_y}{\rho_y^2\Xi^2}(dt - at_kd\phi)^2
 + \rho_y^2\Big(\frac{y'^2}{\Delta_y} + \frac{\theta'^2}{\Delta_\theta}\Big)d\sigma^2\nonumber\\
&\qquad
 + \frac{\Delta_\theta}{\rho_y^2\Xi^2}s_k(ay^2dt - (1+a^2y^2)d\phi)^2
 + (1-ks_k)d\psi^2\Big].
\end{align}

For the internal space $S^5$, the metric of subspace $S^2$ is the same as before 
\begin{equation}
ds_{S^2}^2 = d\varphi^2 + \sin^2\varphi d\chi^2.
\end{equation}
Adding the gauge flux, $\mathcal F = -\kappa\sin\varphi d\varphi\wedge d\chi$,
\begin{align}
g_\text{ind} + \mathcal F
&= \frac1{y^2\rho_y^2\Xi^2}\begin{bmatrix}
 -\Delta_y + \Delta_\theta s_ka^2y^4& 
  \Delta_yat_k - \Delta_\theta s_kay^2(1+a^2y^2)\\
 \Delta_yat_k - \Delta_\theta s_kay^2(1+a^2y^2)
  & -\Delta_ya^2t_k^2 + \Delta_\theta s_k(1+a^2y^2)^2
\end{bmatrix}\nonumber\\
&\quad
\oplus\frac{\rho_y^2}{y^2}\text{diag}\Big[\frac{y'^2}{\Delta_y} + \frac{\theta'^2}{\Delta_\theta},(1-ks_k)\Big]
\oplus\begin{bmatrix}
1& -\kappa\sin\varphi\\ \kappa\sin\varphi& \sin^2\varphi.
\end{bmatrix}.
\end{align}
By calculating the determinant,
\begin{equation}\label{eq:topBH_rot_Lkbi}
\sqrt{-\det(g_\text{ind}+\mathcal F)}
= \frac{\sqrt{1+\kappa^2}\sin\varphi}{y^4\Xi^2}\sqrt{s_k(1-ks_k)}
  \sqrt{\Delta_\theta y'^2 + \Delta_y\theta'^2}
  \Big(1 + (1 - t_k)a^2y^2\Big).
\end{equation}

\subsection{The action and the equation of motions}
The action is from Eq.\ref{eq:topBH_rot_Lkbi} and Eq.\ref{eq:topBH_rot_RR4form},
\begin{align}
S_\text{D5}
&= -T_5\sqrt{1+\kappa^2}\int_\text{AdS}dtd\sigma d\phi d\psi\int_{S^2}d\varphi d\chi\nonumber\\
&\quad
 \sin\varphi\frac{\sqrt{s_k(1-ks_k)}}{\Xi^2}
 \Big(\sqrt{\Delta_\theta y'^2 + \Delta_y\theta'^2}\frac{1+(1-t_k)a^2y^2}{y^4}
  - \tilde\kappa\theta'\Big(\frac1{y^4} + \frac{2a^2}{y^2}(1+t_k)\Big)\Big)\nonumber\\
&= -4\pi(2\pi)^2T_5\Delta t\sqrt{1+\kappa^2}\int d\sigma
 \frac{\sqrt{s_k(1-ks_k)}}{\Xi^2}\nonumber\\
&\qquad
 \Big(\sqrt{\Delta_\theta y'^2 + \Delta_y\theta'^2}
 \Big(\frac1{y^4} + \frac{a^2}{y^2}(1-t_k)\Big)
  - \tilde\kappa\theta'\Big(\frac1{y^4} + \frac{2a^2}{y^2}(1+t_k)\Big)\Big),
\end{align}
where in the second line we performed the integral over $S^2$ and $t$,$\phi$ and $\psi$ on AdS part.
The Lagrangian is 
\begin{align}
\mathcal L 
&= \sqrt{s_k(1-ks_k)}\Big(\sqrt{\Delta_\theta y'^2 + \Delta_y\theta'^2}
  \Big(\frac1{y^4} + \frac{a^2}{y^2}(1-t_k)\Big)
  - \tilde\kappa\theta'\Big(\frac1{y^4} + \frac{2a^2}{y^2}(1+t_k)\Big)\Big)\nonumber\\
&=: S_k(\Phi(y,y',\theta,\theta')X(y,\theta) - \tilde\kappa\theta' Y(y,\theta)).
\end{align}
Here we define the following functions
\begin{subequations}
\begin{align}
\Phi(y,y',\theta,\theta') &:= \sqrt{\Delta_\theta y'^2 + \Delta_y\theta'^2},\\
S_k(\theta) &:= \sqrt{s_k(1-ks_k)},\\
X(y,\theta) &:= \frac1{y^4} + \frac{a^2}{y^2}(1-t_k),\\
Y(y,\theta) &:= \frac1{y^4} + \frac{2a^2}{y^2}(1+t_k).
\end{align}
\end{subequations}
By definition \eqref{eq:topBH_rot_sk}, explicitly
\begin{equation}\label{eq:topBH_rot_Sk}
S_k(\theta) = \begin{cases}
\frac12\sin(2\theta)  & (k=1),\\
\frac12\sinh(2\theta) & (k=-1),\\
1& (k=0). 
\end{cases}
\end{equation}

The derivative for each variables gives
\begin{align}
\frac{\partial\mathcal L}{\partial y'}
&= S_kX\partial_{y'}\Phi,&
\frac{\partial\mathcal L}{\partial y}
&= S_k(X\partial_y\Phi + \Phi\partial_yX 
 - \tilde\kappa\theta'\partial_yY),\\
\frac{\partial\mathcal L}{\partial\theta'}
&= S_k(X\partial_{\theta'}\Phi - \tilde\kappa Y),&
\frac{\partial\mathcal L}{\partial\theta}
&= S_k(X\partial_\theta\Phi + \Phi\partial_\theta X
  - \tilde\kappa\theta'\partial_\theta Y)
 + \partial S_k\frac{\mathcal L}{S_k}.
\end{align}
In the above
\begin{equation}
\partial_{y'}\Phi = \frac{\Delta_\theta y'}\Phi,\;\;
\partial_y\Phi = \frac12\theta'^2\frac{\partial\Delta_y}\Phi,\;\;
\partial_{\theta'}\Phi = \frac{\Delta_y\theta'}\Phi,\;\;
\partial_\theta\Phi = \frac12y'^2\frac{\partial\Delta_\theta}\Phi,
\end{equation}
where $\partial$ in the front of $\Delta_y$ means the derivative by $y$ and that in the front of $\Delta_\theta$ and $S_k$ means the derivative by $\theta$.
The equations of motion are
\begin{subequations}
\begin{align}
\Big(y'S_kX\frac{\Delta_\theta}\Phi\Big)' - S_kX\partial_y\Phi - S_k\Phi\partial_yX + \tilde\kappa\theta'S_k\partial_yY = 0,\\
\Big(\theta'S_kX\frac{\Delta_y}\Phi - \tilde\kappa S_kY\Big)'
- S_kX\partial_\theta\Phi - S_k\Phi\partial_\theta X + \tilde\kappa\theta'S_k\partial_\theta Y 
- \partial S_k\frac{\mathcal L}{S_k} = 0.
\end{align}
\end{subequations}
By separating the second derivative terms, the first equation of motion is 
\begin{equation}
y'' + y'\frac{d}{d\sigma}\log\Big(S_kX\frac{\Delta_\theta}\Phi\Big)
 - \frac12\theta'^2\frac{\partial\Delta_y}{\Delta_\theta}
 - \frac{\Phi^2}{\Delta_\theta}\frac{\partial_yX}X
 + \tilde\kappa\theta'\frac{\Phi}{\Delta_\theta}\frac{\partial_yY}X = 0
\end{equation}
and the second equation of motion is 
\begin{equation}
\theta''
+ \theta'\frac{d}{d\sigma}\log\Big(S_kX\frac{\Delta_y}\Phi\Big)
- \tilde\kappa\frac{y'\partial_yY}X\frac{\Phi}{\Delta_y}
- \frac12y'^2\frac{\partial\Delta_\theta}{\Delta_y} 
- \frac{\Phi^2}{\Delta_y}\frac{\partial_\theta X}X
- \partial\log S_k\frac{\Phi^2}{\Delta_y}
= 0.
\end{equation}

We fixed the gauge $\Phi = 1$.
Let us confirm that this gauge is preserved along the solution.
$\Phi$ changes as
\begin{align*}
\frac{d}{d\sigma}(y'^2\Delta_\theta + \theta'^2\Delta_y)
&= 2y'y''\Delta_\theta + 2\theta'\theta''\Delta_y
 + y'^2\Delta_\theta' + \theta'^2\Delta_y'.
\end{align*}
By the equations of motion, 
\begin{align*}
&\frac{d}{d\sigma}(y'^2\Delta_\theta + \theta'^2\Delta_y)\nonumber\\
&= 2y'\Delta_\theta
  \Big(-y'\frac{d}{d\sigma}\log(S_kX\Delta_\theta)
  + \frac1{\Delta_\theta}\frac{\partial_yX}X
  - \tilde\kappa\frac{\theta'}{\Delta_\theta}\frac{\partial_yY}X
  + \frac12\theta'^2\frac{\partial\Delta_y}{\Delta_\theta}\Big)\\
&\quad
 + 2\theta'\Delta_y
  \Big(-\theta'\frac{d}{d\sigma}\log(S_kX\Delta_y)
  + \frac1{\Delta_y}\frac{\partial_\theta X}X
  + \tilde\kappa\frac{y'}{\Delta_y}\frac{\partial_yY}X
  + \frac12y'^2\frac{\partial\Delta_\theta}{\Delta_y}
  + \frac1{\Delta_y}\partial\log S_k\Big)\nonumber\\
&\quad
 + y'^2\Delta_\theta' + \theta'^2\Delta_y'\\
&= -2\Phi^2\frac{d}{d\sigma}\log(S_kX)
 - 2(y'^2\Delta_\theta' + \theta'^2\Delta_y')
 + 2(\log X)'
 + \theta'^2\Delta_y' + y'^2\Delta_\theta'
 + 2(\log S_k)'\nonumber\\
&\quad
 + y'^2\Delta_\theta' + \theta'^2\Delta_y'\\
&= 2(1-\Phi^2)\frac{d}{d\sigma}\log(S_kX).
\end{align*}
The last expression is zero by using $\Phi = 1$.
Therefore the gauge condition $\Phi=1$ is preserved.

For $\Phi = 1$, the equations of motion are simplified as
\begin{subequations}
\begin{align}
&\frac{d}{d\sigma}(y'S_k\Delta_\theta X)
 - S_k\Big(\frac12\theta'^2\partial\Delta_yX + \partial_yX - \tilde\kappa\theta'\partial_yY\Big) = 0,\\
&\frac{d}{d\sigma}(\theta'S_k\Delta_yX - \tilde\kappa S_kY)
 - S_k\Big(\frac12y'^2\partial\Delta_\theta X + \partial_\theta X - \tilde\kappa\theta'\partial_\theta Y\Big)\nonumber\\
&\hspace{6cm}
  - \partial S_k(X - \tilde\kappa\theta'Y) = 0.
\end{align}
\end{subequations}
Since $Y' = y'\partial_yY + \theta'\partial_\theta Y$, the second equation can be changed as
\begin{equation}
\frac{d}{d\sigma}(\theta'S_k\Delta_yX) 
 - S_k\Big(\frac12y'^2\partial\Delta_\theta X + \partial_\theta X + \tilde\kappa y'\partial_y Y\Big) 
 - \partial S_kX = 0.
\end{equation}
The equation of motions are finally
\begin{subequations}
\begin{align}
&y'' + y'\frac{d}{d\sigma}\log(S_k\Delta_\theta X)
 - \frac1{\Delta_\theta}
  \Big(\frac12\theta'^2\partial\Delta_y + \frac{\partial_yX}X 
 - \tilde\kappa\theta'\frac{\partial_yY}X\Big) = 0,\\
&\theta'' + \theta'\frac{d}{d\sigma}\log(S_k\Delta_yX) 
 - \frac1{\Delta_y}\Big(\frac12y'^2\partial\Delta_\theta 
 + \frac{\partial_\theta X}X
 + \tilde\kappa y'\frac{\partial_yY}X\Big) 
 - \frac{\partial\log S_k}{\Delta_y} = 0.
\end{align}
\end{subequations}

\subsection{Behavior in specific regions}
\paragraph{Zero angular momentum case}
In the case where $a=0$,
$X(y,\theta) = Y(y,\theta) = 1/y^4$,
$\Delta_\theta = \rho_y^2 = \Xi^2 = 1$ and $\Delta_y = 1+ky^2-2my^d = F_k(y)$.
The equations become
\begin{subequations}
\begin{align}
&\text{LHS of $y$ eom}\nonumber\\
&= y'' + y'\frac{d}{d\sigma}\log(S_kX)
 - \Big(\frac12\theta'^2\partial F_k(y) 
 - \frac4y(1 - \tilde\kappa\theta')\Big)\nonumber\\
&= y'' + y'\frac{d}{d\sigma}\log(S_k)
 - \frac12\theta'^2\partial F_k(y) 
 + \frac4y(1 - y'^2 - \tilde\kappa\theta'),\\
&\text{LHS of $\theta$ eom}\nonumber\\
&= \theta'' + \theta'\frac{d}{d\sigma}\log(S_kF_k(y)X) 
 + \frac{4\tilde\kappa y'}{yF_k(y)}
 - \frac{\partial\log S_k}{F_k(y)}\nonumber\\
&= \theta'' + \theta'\frac{d}{d\sigma}\log(S_k)
 + \frac{\theta'y'\partial F_k(y) - 4y'(\theta' - \tilde\kappa)/y - \partial\log S_k}{F_k(y)}.
\end{align}
\end{subequations}
Then these recovers the static case \eqref{eq:topBH_static_eom} with $s_k$ is replaced with $S_k := \sqrt{s_k(1-ks_k)}$.

For the zero flux case, 
\begin{subequations}
\begin{align}
&y'' + y'\frac{d}{d\sigma}\log(S_k)
 - \frac12\theta'^2\partial F_k(y) 
 + \frac4y(1 - y'^2) = 0,\\
&\theta'' + \theta'\frac{d}{d\sigma}\log(S_k)
 + \frac{\theta'y'\partial F_k(y) - 4y'\theta'/y - \partial\log S_k}{F_k(y)}
= 0.
\end{align}
\end{subequations}
The first equation is satisfied by $y=\sigma$ and $\theta=\text{constant}$.
The second equation is satisfied by $\theta = \text{const.}$ as long as $\partial\log S_k = 0$.
This function is (see \eqref{eq:topBH_rot_Sk})
\begin{equation}
\partial\log S_k
= \begin{cases}
2\cot(2\theta)& (k=1),\\
2\coth(2\theta)& (k=-1),\\
0& (k=0).
\end{cases}
\end{equation}
Then for spherical black holes, $\theta = \pi/4$ is a solution, for toroidal black holes, arbitrary constant $\theta$ is a solution and for hyperbolic black holes, there is no such a solution ($\coth x>1$).
These also satisfies the gauge constraint: $\Delta_\theta=1$, $y'=1$, $\theta'=0$ and then $\Phi = 1$.
Summarizing them we find there is a static solution with zero flux for $k=1$ ($\theta = \pi/4$) and $k=0$ ($\theta$ is constant) while there is no solution for $k=-1$ in agreement with the previous case. 

\paragraph{Boundary behavior}
We would like to find the behavior approaching the boundary $y\rightarrow0$.

The factors in the equations behave:
$\partial\Delta_y 
\rightarrow0\; (y\rightarrow0)$ and $\partial\Delta_\theta$ is nonzero keeping $\theta$ dependence.
\begin{align}
\frac{\partial_\theta X}{X}
&= \frac{-\frac{a^2}{y^2}\partial t_k(\theta)}{\frac1{y^4}+\frac{a^2}{y^2}(1-t_k)}
= \frac{-a^2y^2\partial t_k}{1+a^2y^2(1-t_k)} 
\rightarrow0\quad(y\rightarrow0),\\
\frac{\partial_yX}{X}
&= \frac{-\frac4{y^5}-\frac{2a^2}{y^3}(1-t_k)}{\frac1{y^4}+\frac{a^2}{y^2}(1-t_k)}
= \frac{-4/y-2a^2y(1-t_k)}{1+a^2y^2(1-t_k)}
\rightarrow -\frac4y\quad(y\rightarrow0).
\end{align}
The derivatives of $Y$ behaves in the same way. 
The flat AdS solution $x = yX = \kappa y$ with $X = r\cos\theta$ gives
$\cos(\pi/4+\delta\theta(y)) = \kappa y$.
We expect the solution approaches the solution of the flat AdS case on AdS boundary.
In the limit $y\rightarrow0$ and $\delta\theta\rightarrow0$, $\delta\theta' = \kappa y'$.
The gauge constraint becomes $y'^2(\Delta_\theta(\delta\theta=0) + \kappa^2) = 1$.
By solving it for $y'$,
\begin{equation}\label{eq:Rot_toplBH_initial}
y'(\sigma=0) 
= \frac1{\sqrt{\Delta_\theta(\delta\theta=0)+\kappa^2}},\;\;
\delta\theta'(\sigma=0)
= \frac{\kappa}{\sqrt{\Delta_\theta(\delta\theta=0)+\kappa^2}}.
\end{equation}
We impose the initial condition as 
$y(\sigma=0) = 0$, $\delta\theta(\sigma=0) = 0$ and Eq.\eqref{eq:Rot_toplBH_initial} for their first derivatives.

\subsection{Numerical results}
The results are summarized in the following figures.
The fist Figure \ref{fig:topBH_rot0_kappa0_hyperbolic} shows the solution for zero flux and zero angular momentum cases.
As discussed in the previous subsection for a hyperbolic black hole there is not the solution such that the D5-brane exists in the interior of the horizon.

\begin{figure}[h]
\begin{center}
	\includegraphics[width=9cm]{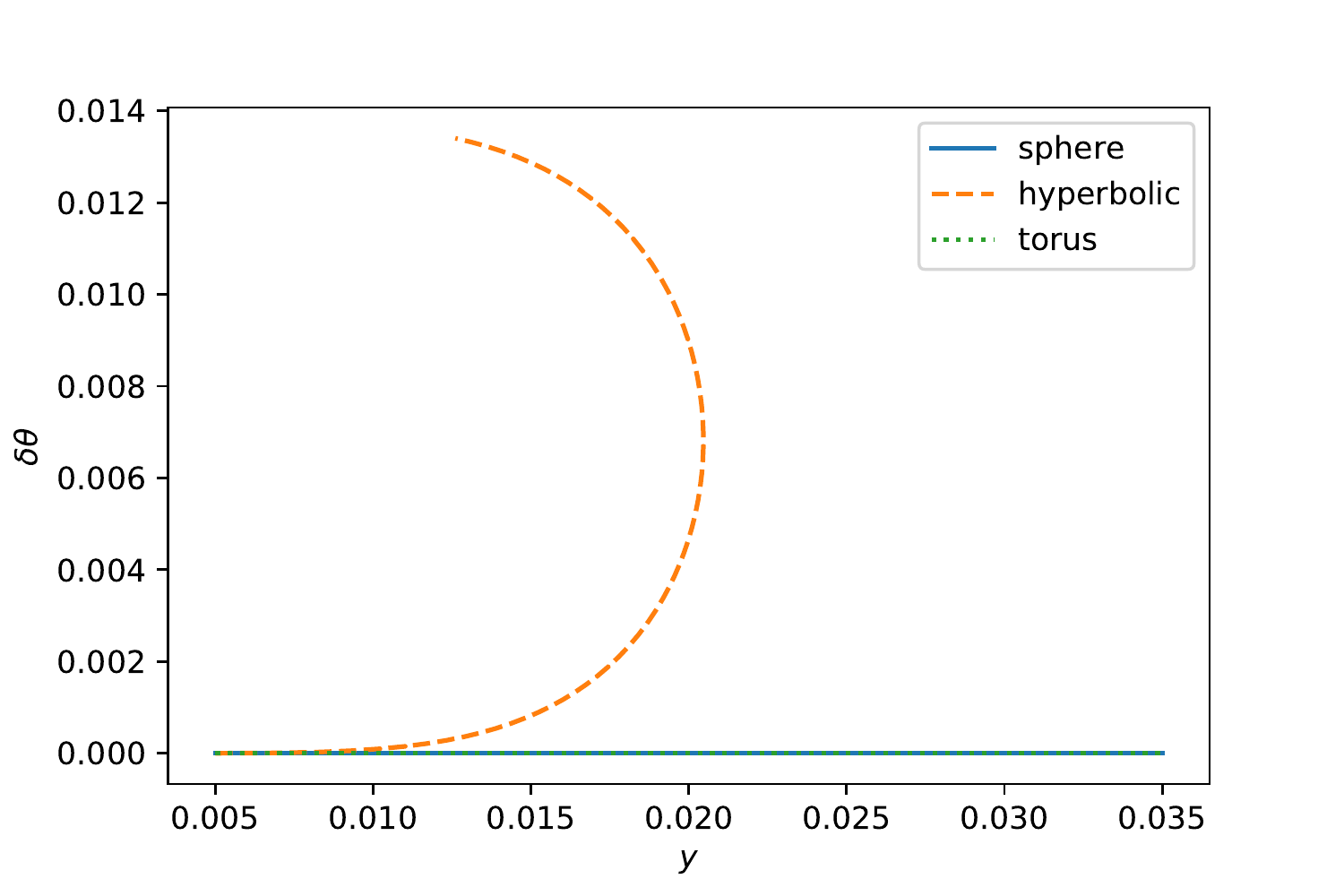}
	\caption{Three different topologies ($\kappa=0$, $a=0$)}
	\label{fig:topBH_rot0_kappa0_hyperbolic}
\end{center}
\end{figure}

\begin{figure}[h]
	\begin{minipage}[t]{0.5\linewidth}
	\includegraphics[width=\linewidth]{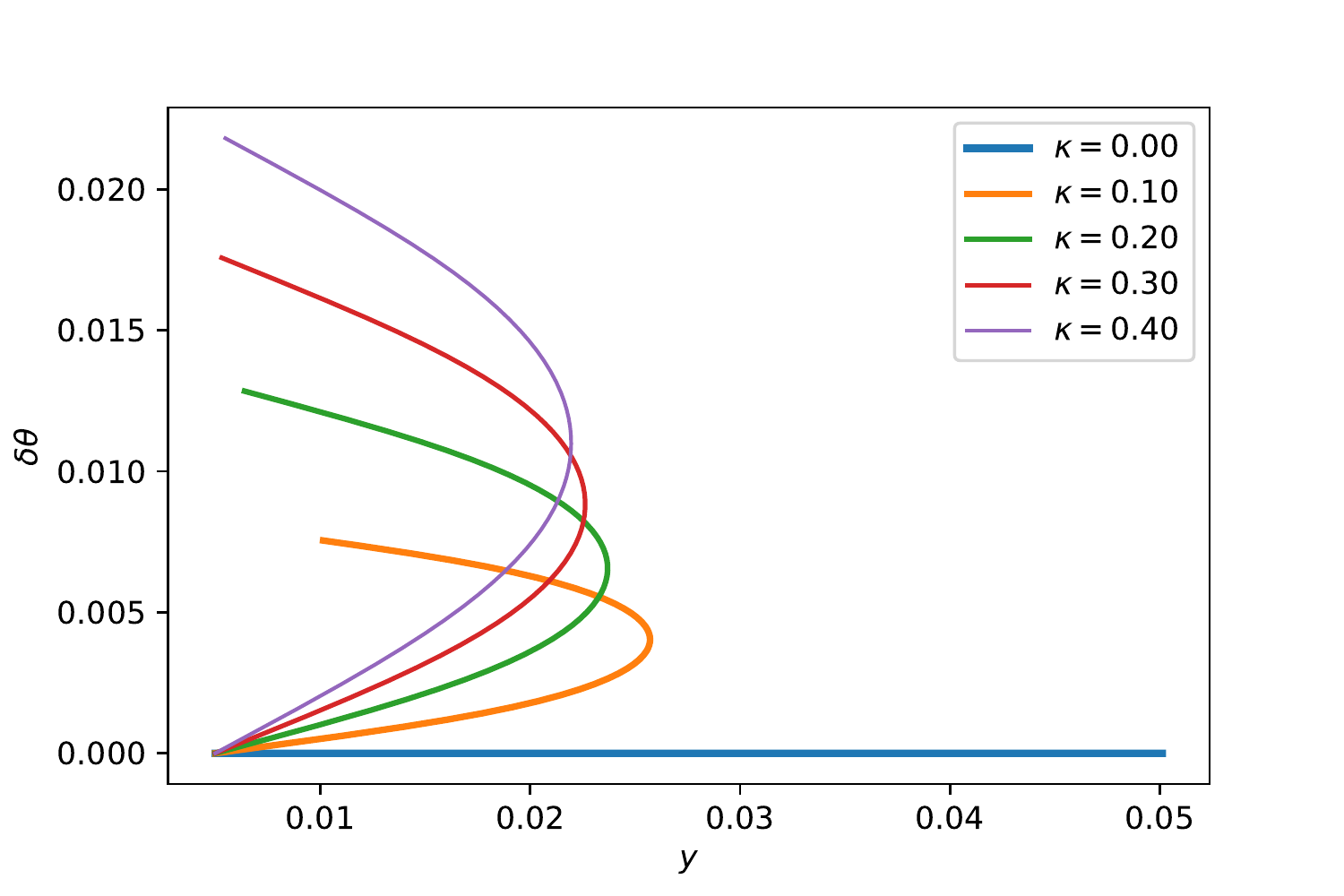}
	\caption{Spherical case $a=0.1$}
	\label{fig:topBH_rot01_kappa_sphere}
	\end{minipage}
\hspace{0.01\linewidth}
 	\begin{minipage}[t]{0.5\linewidth}
 	\includegraphics[width=\linewidth]{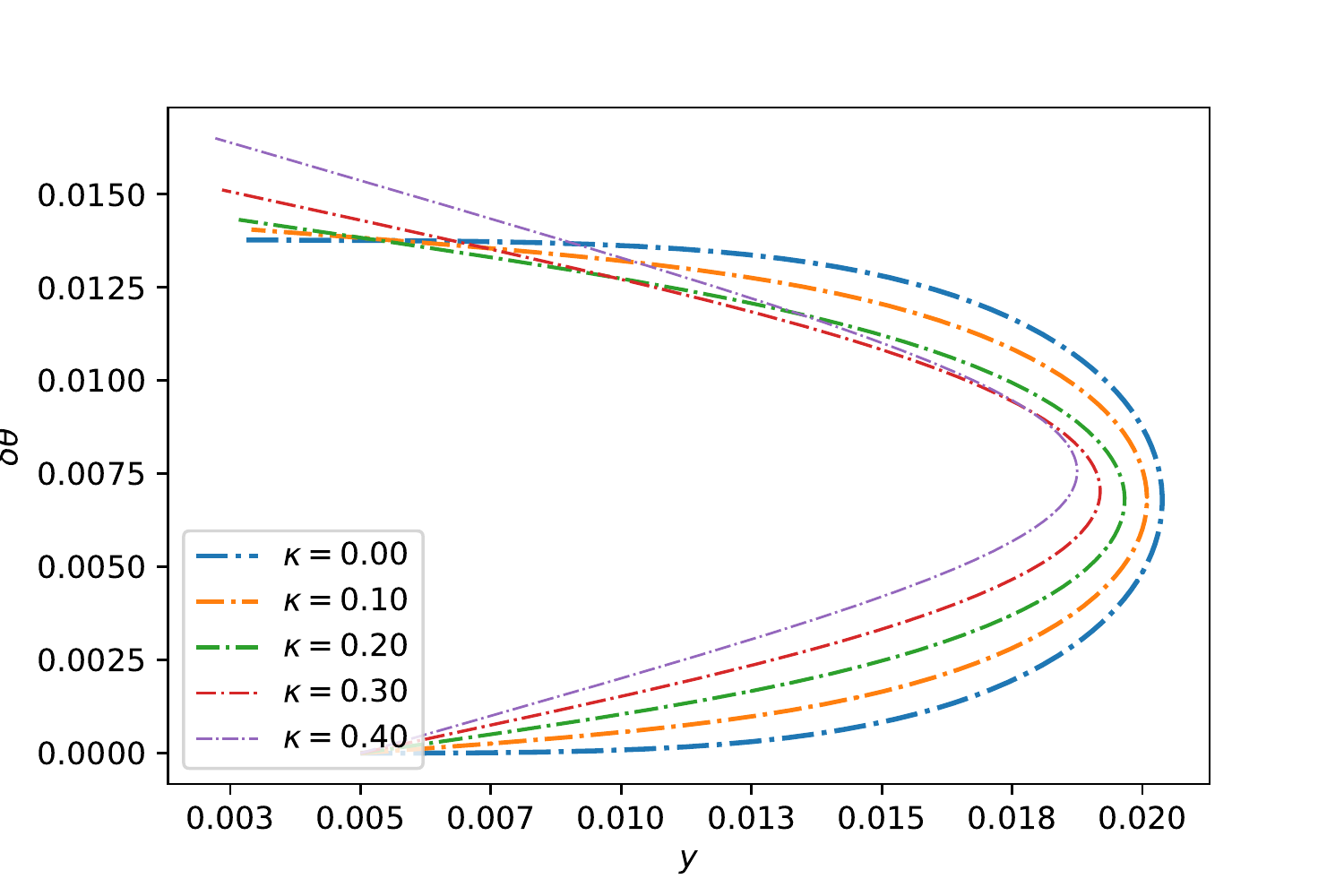}
 	\caption{Hyperbolic case ($a=0.1$)}
 	\label{fig:topBH_rot01_kappa_hyper}
	\end{minipage}
\end{figure}

The following four figures show the flux dependence of the D5-brane. 
Figure \ref{fig:topBH_rot01_kappa_sphere} is the sphere case. 
The black hole mass is fixed to $m=1$ and the angular momentum is fixed to $a=0.1$.
The slope near the AdS boundary changes according to the boundary condition 
$y\approx\kappa\delta\theta|_{\sigma=0}$.
We can see the D5-brane is far apart as the flux value grows.

Figure \ref{fig:topBH_rot01_kappa_hyper} is the hyperbolic case. 
Even the solution begins with zero slope at the boundary, it bends to avoid the horizon.

\begin{figure}[h]
	\begin{minipage}[t]{0.5\linewidth}
	\includegraphics[width=\linewidth]{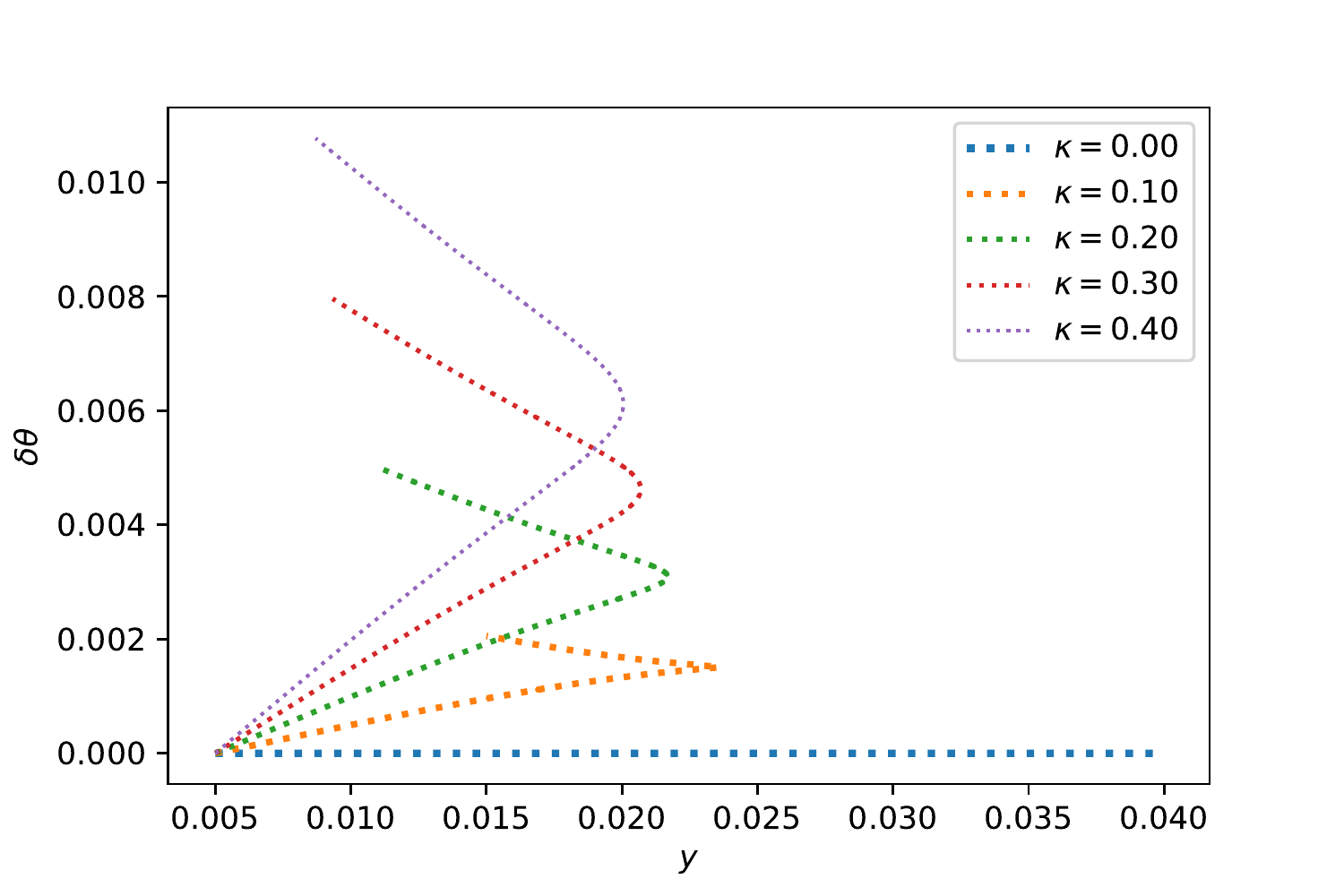}
	\caption{Toroidal case ($a=0.1$)}
	\label{fig:topBH_rot01_kappa_torus_i}
	\end{minipage}
\hspace{0.01\linewidth}
	\begin{minipage}[t]{0.5\linewidth}
	\includegraphics[width=\linewidth]{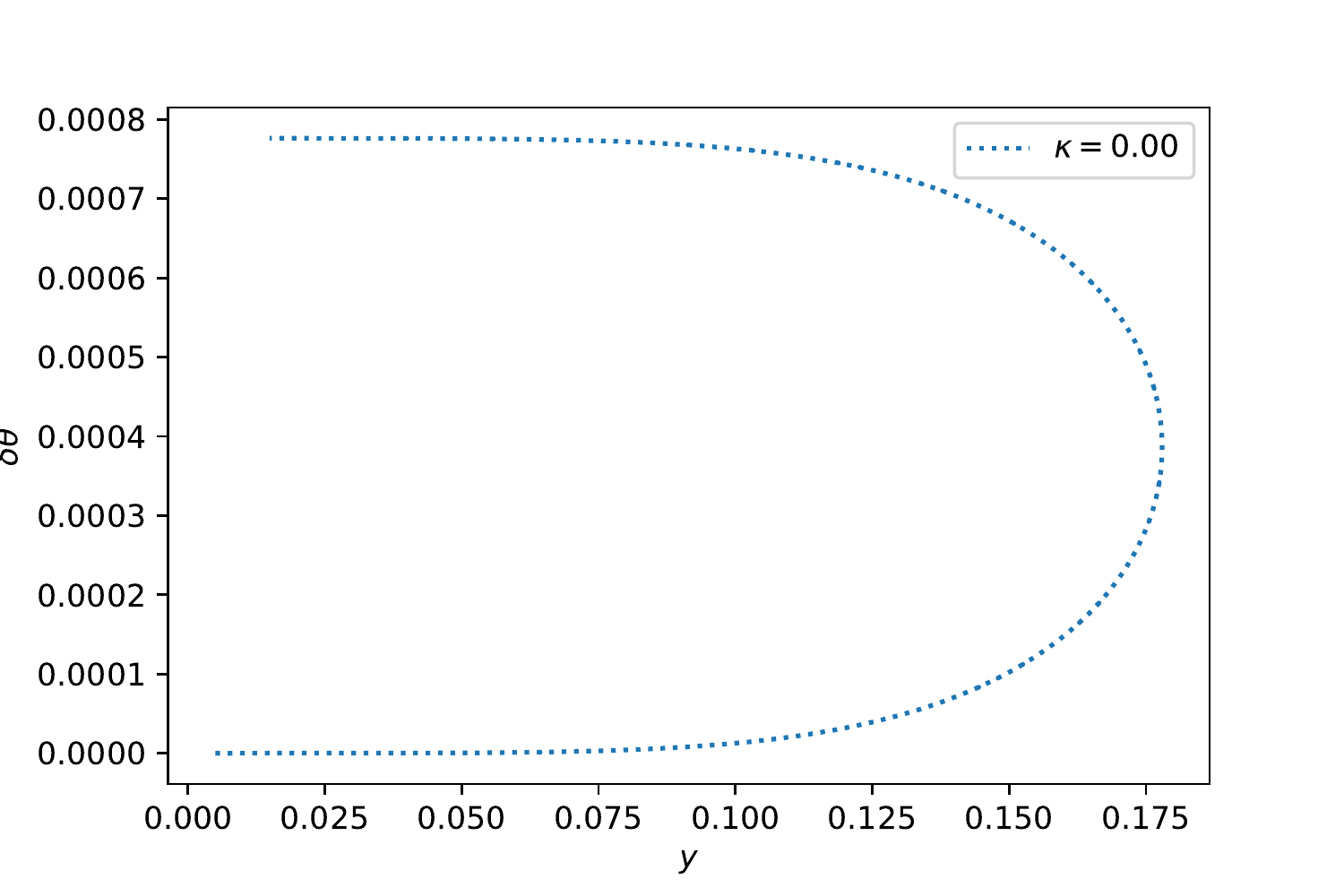}
	\caption{Toroidal case for a broad region ($a=0.1$)}
	\label{fig:topBH_rot01_kappa_torus_ii}
	\end{minipage}
\end{figure}

Figures \ref{fig:topBH_rot01_kappa_torus_i} and \ref{fig:topBH_rot01_kappa_torus_ii} are the torus case.
When the black holes have the angular momentum $a=0.1$, the D5-brane cannot enter the inner of the horizon. 
Since this behavior is not easy to see in Figure \ref{fig:topBH_rot01_kappa_torus_i}, we plot the zero flux solution in a broad region in the next figure (Figure \ref{fig:topBH_rot01_kappa_torus_ii}).

\begin{figure}[h]
	\begin{minipage}[t]{0.5\linewidth}
	\includegraphics[width=\linewidth]{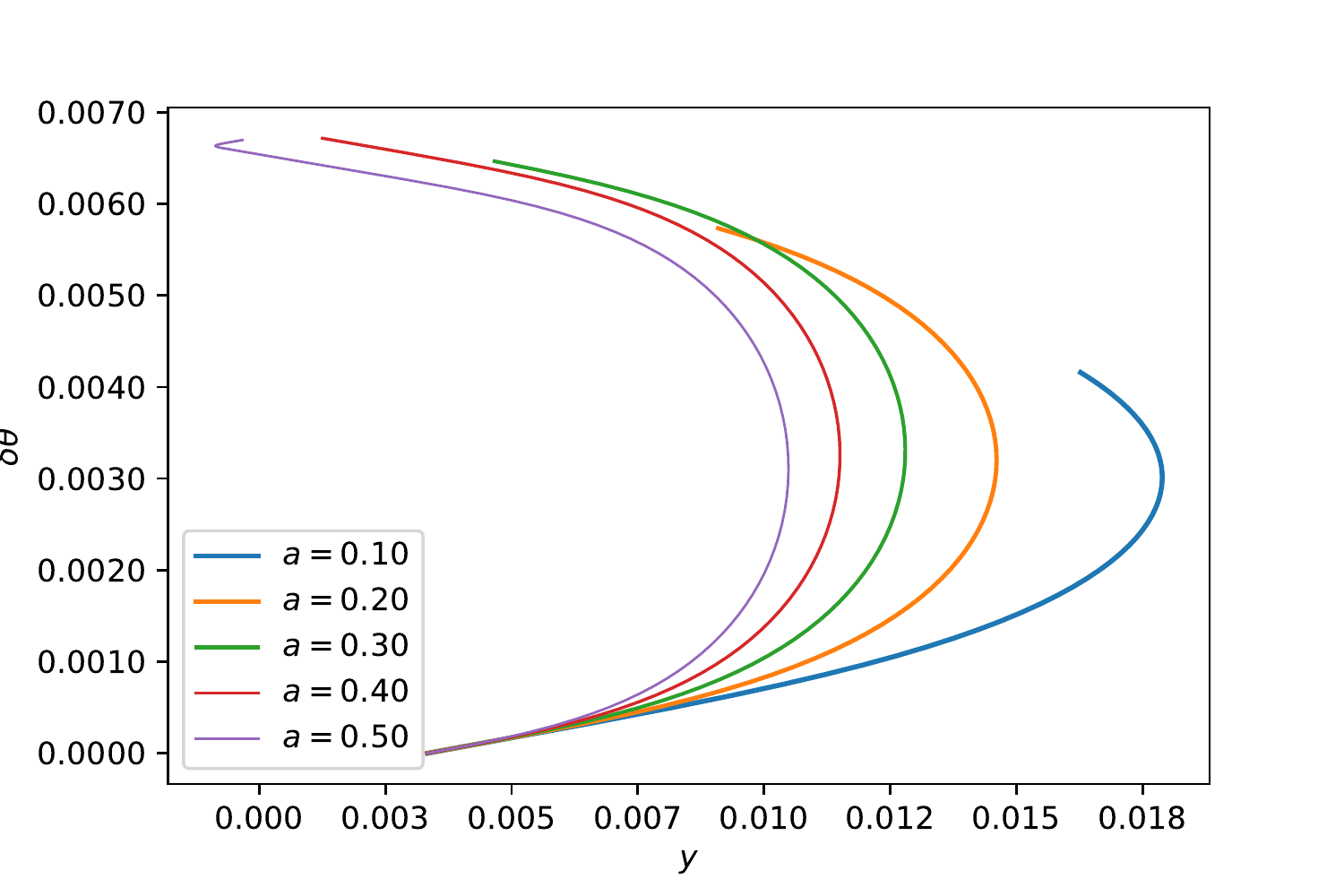}
	\caption{Angular momentum dependence: phere ($\kappa=0.1$)}
	\label{fig:topBH_rot_a_sphere}
	\end{minipage}
\hspace{0.01\linewidth}
	\begin{minipage}[t]{0.5\linewidth}
	\includegraphics[width=\linewidth]{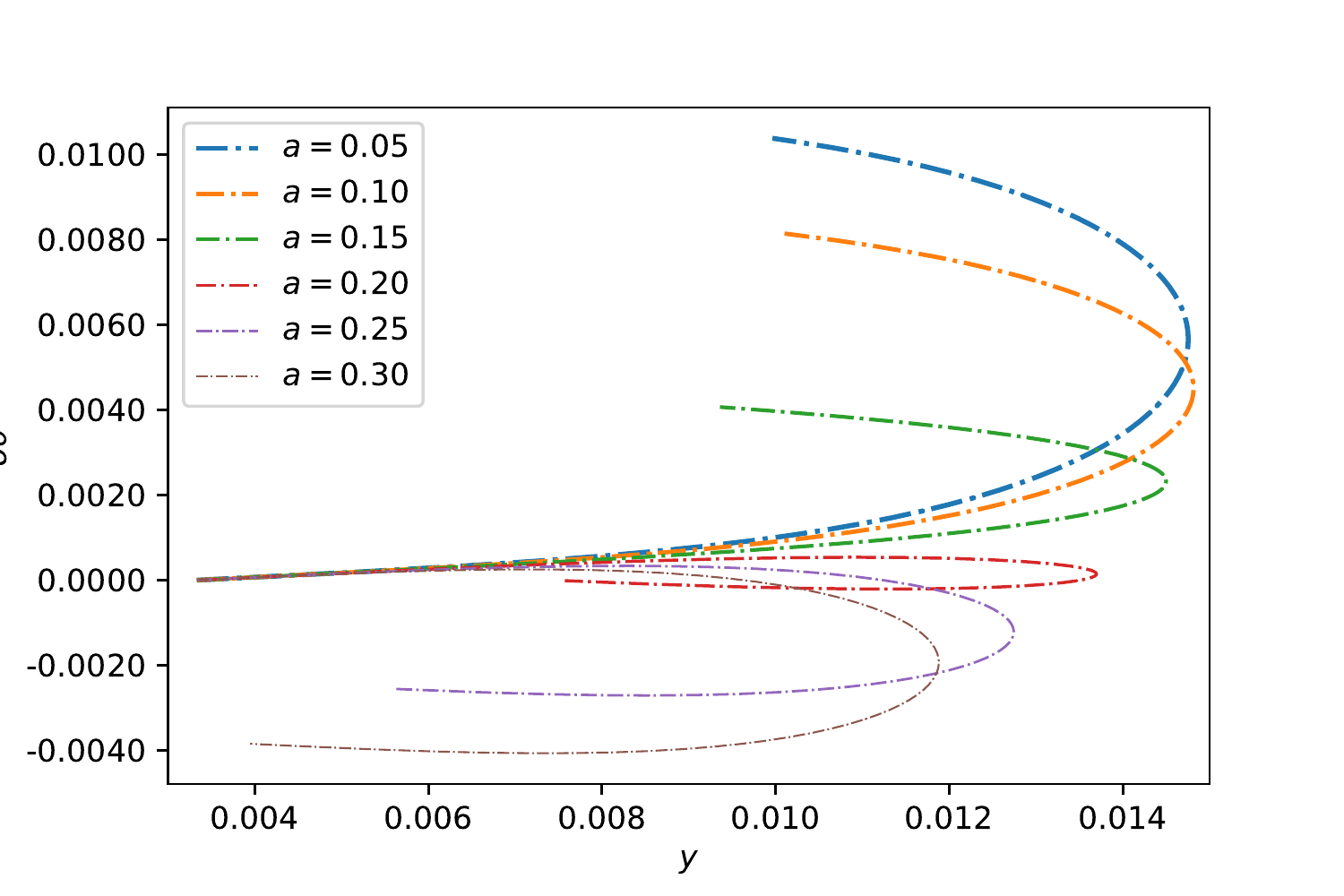}
	\caption{Angular momentum dependence: hyperbolic ($\kappa=0.1$)}
	\label{fig:topBH_rot_a_hyper}
	\end{minipage}
\hspace{0.01\linewidth}
	\begin{minipage}[t]{0.5\linewidth}
	\includegraphics[width=\linewidth]{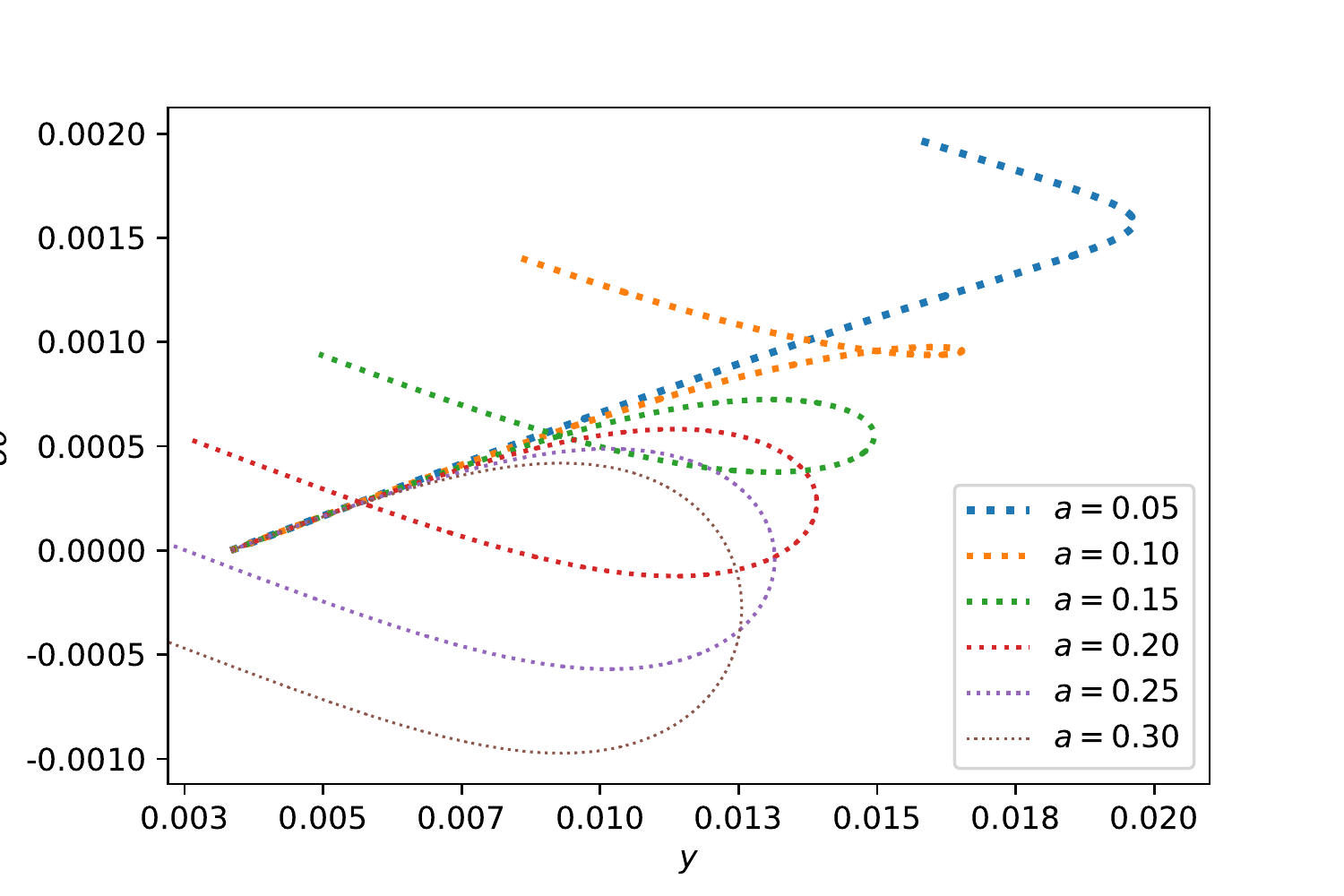}
	\caption{Angular momentum dependence: torus ($\kappa=0.1$)}
	\label{fig:topBH_rot_a_torus}
	\end{minipage}
\hspace{0.01\linewidth}
\end{figure}

Figures \ref{fig:topBH_rot_a_sphere}, \ref{fig:topBH_rot_a_hyper} and \ref{fig:topBH_rot_a_torus} show the angular momentum dependence of the D5-brane solution. 
Since the flux is fixed to $\kappa=0.1$, the slope at the boundary is $\delta\theta=(0.1)\times y$ for all cases.
In all cases the brane departs from the center of the black holes for large angular momentums.
A remarkable feature is that while $\delta\theta$ always increases, that is increases from $\theta=\pi/2$ for sphere case, it can decreases for large angular momentum values for hyperbolic and toroidal black holes.
Therefore in the latter two cases the interface appears in the region $\pi/2<\theta$.

\begin{figure}[h]
\begin{center}
\includegraphics[width=15cm]{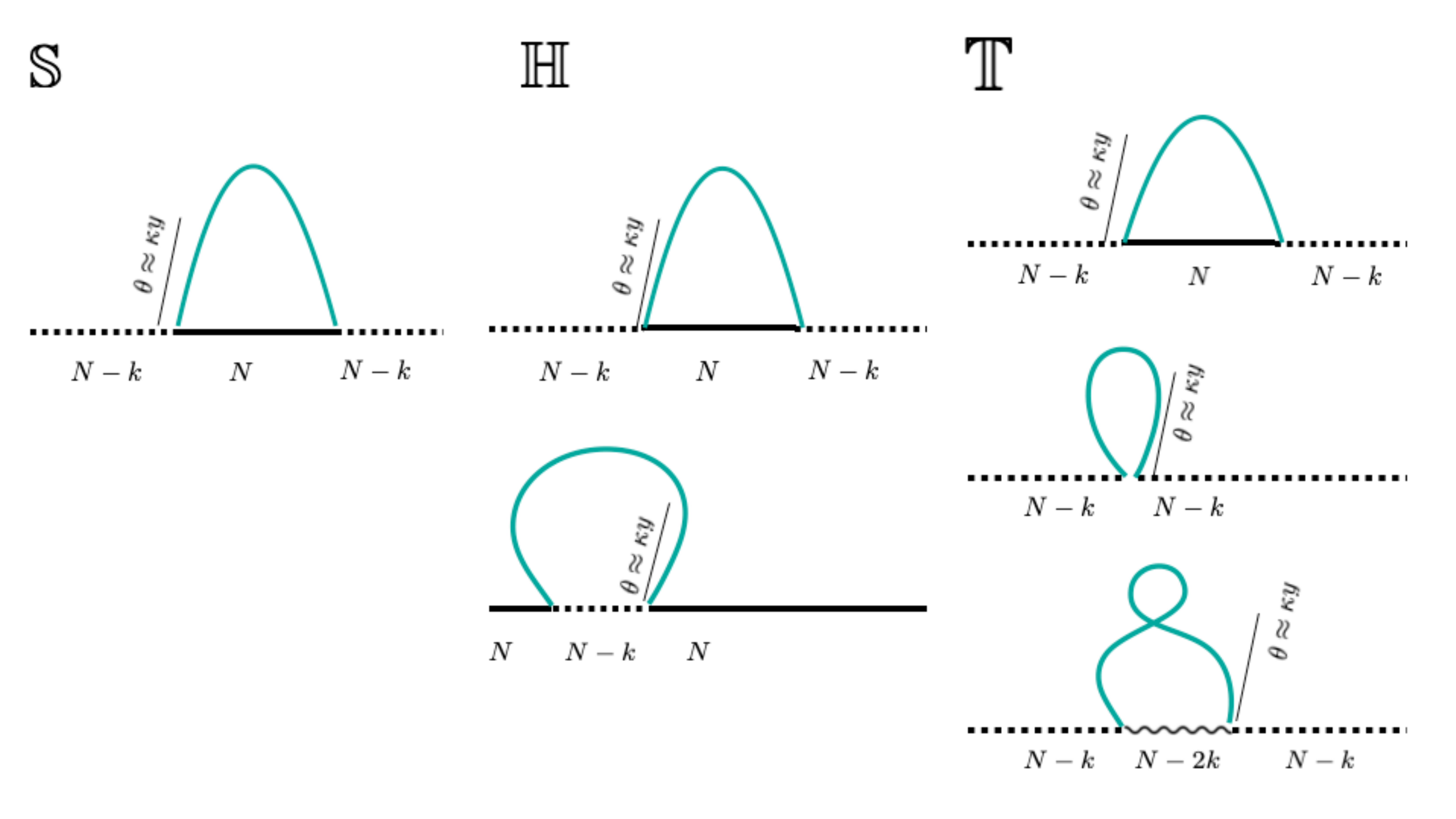}
\caption{Boundary theories for three different topological black holes:
In all cases the slope of the solution near the boundary is $\theta = \kappa y$}
\label{fig:3cases_branesol}
\end{center}
\end{figure}

\newpage
\section{Discussion}\label{sec:Discussion}
In this paper we saw the D5-brane behavior in black hole spacetimes with different topologies. 
These solutions approach to the solution on the flat AdS: $\delta\theta\approx\kappa y$.

 In Section \ref{sec:static_topBHs} we studied the flux dependence of the D5-brane solution for three different kinds of topological black holes. 
As a result, for black holes with hyperbolic structure the system cannot have the trivial solution $\theta = \text{constant}$, while for two other cases (spherical and the toroidal black holes) there are the trivial solution $\theta = \pi/2$ which is the only case the D5-brane can exist in the inner of the horizon. 
Especially for toroidal black holes this constant can be chosen to be arbitrary values. 

In Section \ref{sec:rot_topBHs} we studied the behavior of the D5-branes with small fluxes.
In this section we introduced the black hole angular momentum. 
As in the static black hole cases, this flux value determines the boundary condition of the solution.
We saw that in hyperbolic and the torus case the angular momentum of the black hole changes the behavior of the D5-brane more strongly than sphere case.
The explanation of this behavior is as follows.

We imposed the boundary condition $\delta\theta = \kappa y$ on the AdS boundary. 
As we see in the Figure \ref{fig:topBH_rot01_kappa_sphere},
Figure \ref{fig:topBH_rot01_kappa_hyper} and Figure \ref{fig:topBH_rot01_kappa_torus_i}, the D5-brane can extend towards the region $\theta<\pi/2$ for the hyperbolic and the toroidal cases. 
Especially tor toroidal black holes a self-intersection can occur for sufficient large momentum.

Let us see the boundary gauge theories of the AdS black holes. 
The behavior of the D5-brane is summarized in Figure \ref{fig:3cases_branesol}.
In this figure the vertical direction is the AdS radius $y$ and the horizontal direction represents $\theta$ direction. 

Near the AdS boundary the D5-brane bends towards the side where there are the larger number of D3-brane exist due to the tension of the D3-brane or by the charge condition 
\begin{equation}
k = -\frac{T_5}{T_3}\int\mathcal F
= \frac\kappa{\pi\alpha'}.
\end{equation}

The D5-brane edges are realized as interfaces on boundary.
For the sphere case, the gauge groups of the boundary theories are $SU(N-k)$, $SU(N)$ and $SU(N-k)$ from the left-side to the right-side in the figure.

For the hyperbolic case, solutions with small angular momentum behaves similar to the sphere case.
However, for larger angular momentum the D5-brane can extends the left side ($\delta\theta<0$).
This is the case depicted in lower side of hyperbolic case (Figure \ref{fig:3cases_branesol}).
Following the above rule (D5-branes attract the side where there are the larger number of D3-branes), we identifies the boundary gauge theories have gauge groups $SU(N)$, $SU(N-k)$ and $SU(N)$ in this case. 

For the torus case, a more interesting phenomena occur: there is a self intersection of the D5-brane.
Similar to the sphere case, there are three gauge groups $SU(N-k)$, $SU(N)$ and $SU(N-k)$ from left to right for small angular momentum.
Growing the angular momentum, these two interfaces coincide and the middle side $SU(N)$ collapses. 
Therefore, considering a continuous deformation, for large angular momentums we find the gauge groups are $SU(N-k)$, $SU(N-2k)$ and $SU(N-k)$ from left to right in the figure (bottom of Figure \ref{fig:3cases_branesol}).

\section*{Acknnowledgments}
This research is supported by Department of Physics, Toho University.

\providecommand{\href}[2]{#2}\begingroup\raggedright\endgroup

\end{document}